\documentclass[longauth]{aa} 

\usepackage{graphicx}
\usepackage{longtable}
\usepackage{tabularx}
\usepackage{xltabular}
\usepackage{supertabular}
\usepackage{tabularray}
\usepackage{natbib}
\usepackage{acro}
\usepackage[acronym]{glossaries}
\bibpunct{(}{)}{;}{a}{}{,}
\usepackage{txfonts}
\usepackage{xcolor}
\usepackage{soul}

\usepackage[hyperfootnotes=false]{hyperref}
 \hypersetup{colorlinks=true,allcolors=[rgb]{0,0,1}}
 \defcitealias{Vieuville_2019}{DLV19}
 \defcitealias{Richard_2021}{R21}

\begin{document} 
   \title{Probing the faint end Luminosity Function of Lyman Alpha Emitters at
3<$z$<7 behind 17 MUSE lensing clusters
   }
   \titlerunning{Faint end of LAEs LF}
\authorrunning{Thai et al.}

   \author{T. T. Thai$^{1,2,3}$\thanks{tranthai.physics@gmail.com}, P. Tuan-Anh$^{2,3}$\thanks{ptanh@vnsc.org.vn}, R. Pello$^{1}$\thanks{roser.pello@lam.fr}, I. Goovaerts$^{1}$, J. Richard$^{4}$, A. Claeyssens$^{5}$, G. Mahler$^{6,7}$, D. Lagattuta$^{6,7}$, G. de la Vieuville$^{10}$, E. Salvador-Sol\'{e}$^{8}$, T. Garel$^{9,4}$, F.E. Bauer$^{11,12,13}$, A. Jeanneau$^{4}$, B. Cl\'{e}ment$^{12}$, J. Matthee$^{14}$ }       

   \institute{Aix Marseille Universit\'{e}, CNRS, CNES, LAM (Laboratoire d’Astrophysique de Marseille), UMR 7326, IPhU, 13388 Marseille, France
         \and
         Department of Astrophysics, Vietnam National Space Center, Vietnam Academy of Science and Technology, 18 Hoang Quoc Viet, Hanoi, Vietnam
         \and
         Graduate University of Science and Technology, VAST, 18 Hoang Quoc Viet, Cau Giay, Vietnam
         \and
         Univ Lyon, Univ Lyon1, Ens de Lyon, CNRS, Centre de Recherche Astrophysique de Lyon UMR5574, F-69230, Saint-Genis-Laval, France
        \and
        Department of Astronomy, Oskar Klein Centre, Stockholm University, AlbaNova University Centre, SE-106 91 Stockholm, Sweden
        \and
        Institute for Computational Cosmology, Durham University, South Road, Durham DH1 3LE, UK
        \and
        Centre for Extragalactic Astronomy, Durham University, South Road, Durham DH1 3LE, UK
        \and
       Departament de F\'{i}sica Qu\`{a}ntica i Astrof\'{i}sica, Institut de Ci\`{e}ncies del Cosmos. Universitat de Barcelona, E-08028 Barcelona, Spain   
      \and
      Observatoire de Gen\`{e}ve, Université de Gen\`{e}ve, 51 Ch. des Maillettes, 1290 Versoix, Switzerland
      \and
      7 Avenue Cuvier, F-78600, Maisons-Laffitte, France
      \and
      Instituto de Astrof\'{i}sica and Centro de Astroingenier\'{i}a, Facultad de F\'{i}sica, Pontificia Universidad Cat\'{o}lica de Chile, Casilla 306,Santiago 22, Chile
      \and
      Millennium Institute of Astrophysics, Nuncio Monse\~{n}or S\'{o}tero Sanz 100, Of 104, Providencia, Santiago, Chile
      \and
      Space Science Institute, 4750 Walnut Street, Suite 205, Boulder, Colorado 80301, USA
      \and
      Department of Physics, ETH Zurich, Wolfgang-Pauli-Strasse 27, CH-8093 Zurich, Switzerland
         }

   \date{Received xxx 2023; accepted }

 
  \abstract
   { This paper presents a study of the galaxy Lyman-alpha luminosity function (LF) using a large sample of 17 lensing clusters observed by the Multi-Unit Spectroscopic Explorer (MUSE) at the ESO Very Large Telescope (VLT). Magnification from strong gravitational lensing by clusters of galaxies and MUSE spectroscopic capabilities allow us to blindly detect LAEs without any photometric pre-selection, reaching the faint luminosity  regime.} 
   {The present work aims to constrain the abundance of Lyman-alpha Emitters (LAEs) and quantify their contribution to the total cosmic reionization budget.}
    {600 lensed LAEs were selected behind these clusters in the redshift range 2.9 < $z$ < 6.7, covering four orders of magnitude in magnification-corrected Ly-$\alpha$ luminosity (39.0<\text{log }($L$)[\text{erg s}$^{-1}$]<43.0). 
   These data are collected behind lensing clusters meaning an increased complexity in the computation of the LF to properly account for magnification and dilution effects.  We apply non-parametric $V_\text {max}$ method to compute the LF, by carefully determining the survey volume where an individual source could have been detected. The method used in this work follows the recipes originally developed by \cite {Vieuville_2019} (hereafter \citetalias{Vieuville_2019}) with some improvements to better account for the effects of lensing when computing the effective volume. }
       {The total co-moving volume at $2.9<z<6.7 $ in the present survey is $\sim$50 10$^{3}$ Mpc$^3$. Our LF points in the bright end (log($L$) [\text{erg s}$^{-1}$] > 42) are consistent with those obtained from blank field observations. In the faint luminosity regime, the density of sources is well described by a steep slope, $\alpha \sim\ -2$ for the global redshift range. This value is consistent with those from \cite{Herenz_2019} and is $20\%$ steeper than that of \citetalias{Vieuville_2019}. Up to log($L$) [\text{erg s}$^{-1}$] $\sim$ 41, the steepening of the faint end slope with redshift, suggested by the earlier work of \citetalias{Vieuville_2019} is observed, but the uncertainties remain large. A significant flattening is observed towards the faintest end, for the highest redshift bins (log($L$)[\text{erg s}$^{-1}$] $<$ 41).}
   {
   When taken at face values, the steep slope at the faint-end causes the SFRD to dramatically increase with redshift, implying that LAEs could play a major role in the process of cosmic reionization. The flattening observed towards the faint end for the highest redshift bins still needs further investigation. This turnover is similar to the one observed for the UV LF at $z$ $\ge$ 6 in lensing clusters, with the same conclusions regarding the reliability of current results \citep[e.g.][]{Atek_2018, Bouwens_2022}. Improving the statistical significance of the sample in this low-luminosity high-redshift regime is a difficult endeavour susceptible to yield invaluable leads for understanding the process of reionization.
   }


   \maketitle
%

\section{Introduction}
\label{sec: 1}

A few hundred thousand years after the Big Bang, the Universe entered the Dark Ages; there were no light sources other than the cosmic microwave background radiation. The temperature of the Universe cooled down enough for electrons and protons to combine to form neutral hydrogen atoms. Objects started to collapse under gravity to form the first stars and galaxies. The photons from these first structures, at redshift range $6<z<12$, started to ionize the surrounding environments. The Universe entered the cosmic reionization era and was fully ionized at redshift $z\sim6$ according to observations of the Gunn Peterson trough from quasar spectra \citep{Fan_2006, Goto_2006}.

Hydrogen is the most abundant element in the Universe, and the Ly-$\alpha$ transition is a major tool. 
\cite{Partridge_1967} predicted the existence of this transition in high-$z$ galaxies, Lyman alpha emitters (LAEs), and this was confirmed in the middle of 1990s using both large (8-10m) ground based telescopes and the Hubble Space Telescope (HST). Since then, the Lyman-$\alpha$ emission has become one of the key tools to explore the Universe at the cosmic reionization stage. LAEs have been found at a wide range of redshifts, up to $z\sim 9$  \citep[see e.g.][]{Oesch_2015, Claeyssens_2022, Vieuville_2019, Richard_2021}.

The main sources responsible for cosmic reionization are still being debated to date. It is evident that quasars are unlikely to play an important role \citep[see][for example]{Willott_2010,McGreer_2013}. The star forming galaxies, such as LAEs and Lyman Break Galaxies (LBGs), are the more likely sources based on numerous observations \citep[see e.g.][]{Robertson_2013,Robertson_2015,Bouwens_2015a}. Their contribution to cosmic reionization is quantified via the star formation rate density (SFRD). This in turn strongly depends on the shape of the galaxy luminosity function, particularly at the faint end, and on the escape fraction of ionizing photons that mainly relates to HI opacity, covering factor, dust content, and geometrical considerations  \citep{Giavalisco_1996, Kunth_1998, Hayes_2014}. 

The galaxy Luminosity Function (hereafter LF) is the number of galaxies in a given luminosity range per unit co-moving volume. A popular parametric form of the function was proposed by \cite{Schechter_1976}, and has been used extensively since then. The LF also puts a strong constraint on the theoretical models of galaxies formation \citep{Mao_2017, Kobayashi_2007}. The Schechter LF has three parameters: $\Phi *$, the normalization factor, $L*$, the characteristic luminosity defining the transition between the exponential part of the function (at bright luminosity) and the power law part (at faint luminosity) and $\alpha$, the faint end slope. All three parameters have been studied extensively for LAEs. Fixing the faint end slope at the fiducial values of $-1.8$ at $z$=0.5 and $-2.5$ at $z$=7, the evolution of the other two parameters ($\Phi *$, $L*$) as a  function of redshift can be roughly summarized as follows: they rise up in the range 0 to 3 \citep{Deharveng_2008}, moderate/no evolution from $z\sim 3$ to $z\sim 6$ \citep{Ouchi_2008}, and then drop down beyond $z\sim 6$ \citep{Kashikawa_2011}.

Much progress has been made studying the LAE LF in the bright luminosity regime, i.e., log($L$) [\text{erg s}$^{-1}$] > 42 \citep[see a recent review by][]{Ouchi_2020}. \citet{Spinoso_2020} studied hyper LAEs (log ($L$) [\text{erg s}$^{-1}$] > 10$^{43.3}$) in redshift range $2.2<z<3.3$ from Javalambre Photometric Local Universe Survey (J-PLUS), reporting a Schechter slope of $\alpha$=$-1.35$\,$\pm\,$0.84. Using a sample of 166 LAEs at $z\sim$3.1 obtained from Subaru Suprime-Cam, covering luminosity range of log ($L$)[erg s$^{-1}$]= 42-43.5, \cite{Guo_2020} found log$_{10}$($\phi*$)=$-3.30^{+0.09}_{-0.1}$, log($L*$)=42.91$^{+0.13}_{-0.14}$ while the slope value of $\alpha$ was fixed at $-$1.6. At higher redshifts, \cite{Konno_2018} constructed the LF from 1266 LAEs for redshifts $z=5.7$ and $z=6.6$ obtained from Subaru/Hyper Suprime-Cam survey, with luminosities in the range 10$^{42.8-43.8}$ erg s$^{-1}$. They measured steep slopes of $\alpha$=$-2.6^{+0.6}_{-0.4}$ and $-2.5^{+0.5}_{-0.5}$ at $z=5.7$ and $z=6.6$ respectively.  \cite{Itoh_2018} studied 34 LAEs also observed by the Subaru Telescope at even higher redshift, $z\sim$ 7.0, covering a sky area of 3.1 deg$^2$. However, in most studies the steep faint end slope is considered as a fixed parameter due to the lack of sources observed at faint luminosities, for instance $\alpha=-2.5$ in this case. This value was also assumed by \cite{Konno_2014}  to study LAEs at redshift $z\sim7.3$. More data at faint luminosities are necessary to constrain the faint-end slope.

Recently, 3D/IFU spectroscopy in pencil beam mode, obtained with the Multi-Unit Spectroscopic Explorer (MUSE) of the ESO Very Large Telescope (VLT) has allowed to detect faint LAEs in the redshift range of 2.9 < $z$ < 6.7. With a sample of 604 LAEs probing down to a luminosity of log($L$) [\text{erg s}$^{-1}$] <41.5 obtained from the MUSE Hubble Ultra Deep Field Survey, \cite{Drake_2017} measured faint end slope values of $-2.03^{+1.42}_{-0.07}$ at $z \sim 3.44$ and $-2.86^{+0.76}_{-\infty}$ at $z\sim 5.48$. The MUSE-Wide survey probed 237 LAEs with luminosities in the range 42.2<log($L$) [\text{erg s}$^{-1}$] <43.5 \citep{Herenz_2019}. They measured the three parameters: $\alpha=-1.84^{+0.42}_{-0.41}$, log$ L*$[erg s$^{-1}$]=$42.2^{+0.22}_{-0.16}$ and log$\Phi*[\text{Mpc}^{-3}]$=-2.71, and found no evolution of the LF in the explored redshift range of 2.9 < $z$ <6.7.

The Lensed Lyman-Alpha MUSE Arcs Sample (LLAMAS) project, which is part of MUSE Lensing Cluster Survey (PI: J. Richard, \citealt{Richard_2021}, \citealt{Claeyssens_2022}), helps to reach the fainter luminosity regime, thanks to magnification from strong gravitational lensing. \citetalias{Vieuville_2019} studied a sample of four lensing clusters, including 156 LAEs with luminosities in the range  39< log($L$)[\text{erg s}$^{-1}$] < 43. The present work extends the previous study to a sample of 17 clusters presented in \citet{Claeyssens_2022}, probing the same luminosity range but with $\sim$4 times more sources, $\sim$600 LAEs. With this unprecedented dataset, we expect to set strong constraints on the faint-end shape of the LF, as well as its evolution with redshift, and hence on the contribution of LAE population to cosmic reionization.

The structure of the paper is as follows. In Sect. \ref{sec: 2}, we briefly describe the MUSE cubes used, together with their ancillary HST data. Mass models for lensing clusters and flux measurements of LAEs are presented in Sect. \ref{sec: 3}. We summarize the method used for the construction of the LF in Sect. \ref{sec: 4}. Results of LF fitting are presented in Sect. \ref{sec: 5}. Our results in comparison with previous ones and the implications for cosmic reionization are discussed in Sect. \ref{sec: 6}. Conclusions are given in Sect. \ref{sec: 7}.

Throughout the paper, we adopt the following cosmology parameters $ \Omega_m$ = 0.3 and $\Omega_{\lambda}$ = 0.7 , with $H _0=70$ km s$^{-1}$ Mpc$^{-1}$.

\section{Data}\label{sec: 2}
    \subsection{MUSE data cube}\label{subsec: 2.1}

The Multi Unit Spectroscopic Explorer (MUSE) is installed on the Yepun telescope of the ESO Very Large Telescope (VLT) \citep{Bacon_2010}. It has a spectral resolution of R$\sim $3000 and a Field of View (FOV) of $\sim 1'\times 1'$  in Wide Field Mode (WFM). The MUSE datacubes in the present work have two sky plane coordinates and one wavelength coordinate. The sky plane coordinates have $\sim 300\times 300$ spatial pixels each measuring $0.2\times 0.2$ arcsec$^2$. The wavelength coordinate has 3681 channels ranging from 4750 $\AA$ to 9350 $\AA$ effectively detecting Lyman-$\alpha$ emission at redshifts of $2.9 < z <6.7$.

The data reduction for the first sample of twelve lensing clusters is described in \cite{Richard_2021} (hereafter \citetalias{Richard_2021}). They made use of the recipes by \cite{Weilbacher_2020} with some improvements for crowded fields such as lensing clusters. The same procedure was applied for the five new datacubes (MACS0451, MACS0520, A2390, A2667 and AS1063). These clusters have been observed both with Adaptive Optics and in standard modes (WFM-NOAO-N), except for the BULLET cluster (WFM-NOAO-E), which has a spectral range slightly extended towards the ultraviolet. These observations are well centered on the core regions of the clusters to maximize the chances to detect strong emissions of LAEs. Seeing values, i.e. the FWHM of the Point Spread Function at 700nm, vary from 0.52 to 1.02 arcsec among clusters of the sample. Integration time varies from 2 to 14 hours. Detailed information of the 17 clusters used in this work is shown in Table \ref{table: cluster_info}.

  \begin{table*}
     \caption{Information of 17 lensing clusters (18 fields) observed by MUSE/VLT.}
     $$     
        \begin{array}{p{0.1\linewidth}p{0.1\linewidth} p{0.1\linewidth}p{0.04\linewidth}p{0.13\linewidth}p{0.11\linewidth}p{0.04\linewidth}p{0.09\linewidth}p{0.07\linewidth}}
            \hline
            \hline
            \noalign{\smallskip}
       Cluster & RA & DEC & z & Programme ID& Notes & Seeing value & MUSE depth (hrs) & N. LAEs\\
        \noalign{\smallskip}
        \hline
        A2390 & 21:53:36.823   &+17:41:43.59     & 0.228     & 094.A-0115  & &  0.75    & 2     &7\\
        A2667 & 23:52:28.400   &-6:05:08.00     & 0.233     &  094.A-0115 & & 0.62    & 2      &14\\
        A2744 & 00:14:20.702  &-30:24:00.63    &   0.308   &  094.A-0115, 095.A-0181, 096.A-0496 &  MACS, FF& 0.61    &3.5-7  &128\\ 
        A370  & 02:29:53.122    &-01:34:56.14    &0.375& 094.A-0115, 096.A-0710& FF& 0.66    &   1.5-8.5 &41\\
        AS1063  &22:48:43.975   &-44:31:51.16
    &    0.348& 60.A-9345, 095.A-0653 & FF &  1.02    &3.9 &20\\
        BULLET  &06:58:38.126 &-55:57:25.87    &   0.296   &094.A-0115& &0.56    &2   &  11\\
        MACS0257&   02:57:41.070 &-22:09:17.70&    0.322& 099.A-0292, 0100.A-0249,  0103.A-0157& MACS&  0.52&   8&  24\\
        MACS0329&   03:29:41.568 &-02:11:46.41    &   0.450   &   096.A-0105  & MACS, CLASH&  0.69    &2.5    &16\\
        MACS0416N   &04:16:09.144   &-24:04:02.95   &0.397  &094.A-0115, 0100.A-0763 &MACS, CLASH, FF   &0.53   &17 &45\\
        MACS0416S   &04:16:09.144   &-24:04:02.95   &0.397  &094.A-0525& MACS, CLASH, FF &0.65   &11-15  &32  \\
        MACS0451    &04:51:54.647   &+00:06:18.21   &0.430& 098.A-0502, 0104.A-0489, 106.218M  & MACS     &0.58    &8  &21  \\
        MACS0520    &05:20:42.046   &-13:28:47.58   &0.336  & 098.A-0502, 0104.A-0489, 106.218M & MACS &0.57   &8  &19 \\
        MACS0940 &09:40:53.698   &+07:44:25.31   &0.335  &098.A-0502, 0101.A-0506& MACS    &0.571  &8  &48         \\
        MACS1206  &12:06:12.149   &-08:48:03.37   &0.438  &095.A-0181,  097.A-0269 & MACS, CLASH   &0.521  &4-9    &49 \\
        MACS2214    &22:14:57.292   &-14:00:12.91   &0.502  &099.A-0292, 0101.A-0506, 0103.A-0157, 0104.A-0489 & MACS &0.55   &7  &17    \\
        RXJ1347 &13:47:30.617   &-11:45:09.51   &0.451  &095.A-0525, 097.A-0909&MACS, CLASH &0.551      &2-3    &72     \\
        SMACS2031   &20:31:53.256   &-40:37:30.79   &0.331  &60.A-9100&MACS  &0.79   &10 &20      \\
       SMACS2131   &21:31:04.831   & -40:19:20.92   &0.442  &0101.A-0506, 0102.A-0135, 0103.A-0157&MACS  &0.59   &7  &16       \\ 
       \hline
                    &         &      &\textbf{Total:}      &   &                                       &       &\textbf{107-128}&\textbf{600}\\
                    \hline     
            \noalign{\smallskip}
        \end{array}   
  $$
   \label{table: cluster_info}
  \end{table*}

\subsection{Lensing Clusters sample}
The redshifts of the lensing clusters range between 0.2 $<z_{cl}<$ 0.6. These are efficient to detect LAEs as background sources, with redshifts in the range 2.9 < $z$ < 6.7. These data are part of the Lensed Lyman-Alpha MUSE Arcs Sample (LLAMAS) project, with LAEs selected from MUSE and HST observations. HST observations of each lensing cluster have at least one high resolution broad band filter available to warrant an overlapping with MUSE spectral range. The preliminary mass models of these clusters were built based on HST observations, and were particularly useful to guide the location of the MUSE fields relatively to the critical lines and the regions hosting multiple images. The reader should refer to \citetalias{Richard_2021}, \cite{Claeyssens_2022} and Sect. \ref{sec: 3} 

for the updated and detailed lensing models adopted for these clusters. 

The present sample includes 17 cluster fields with 18 datacubes observed by the MUSE GTO program. Three of them: A2390, A2667, and A2744, were studied before \citetalias{Vieuville_2019} but have since been re-reduced to have better quality data in terms of S/N. Three clusters, A2744, A370 and RXJ1347, are observed in mosaic mode with $2\times2$ MUSE FoVs, and MACS1206 is observed with $3\times1$ MUSE FoVs. The Northern part of MACS0416 is observed by an ESO program 0100.A-0764 (PI: Vanzella), which is then combined with the GTO observation to produce an improved datacube in terms of exposure time. 

As seen below, with the present sample of lensed LAEs, it is possible to construct a more robust LF than the previous work of \citetalias{Vieuville_2019}. Our sample probes the same four orders of magnitude in Lyman alpha luminosity, down to 10$^{39}$ erg s$^{-1}$, however we significantly improve the galaxy luminosity distribution in terms of statistics and provide a better coverage at the faint end. Information on these clusters and the number of LAEs behind them is shown in Table \ref{table: cluster_info}.

\begin{figure*}
   \centering   
    \includegraphics[height=16cm]{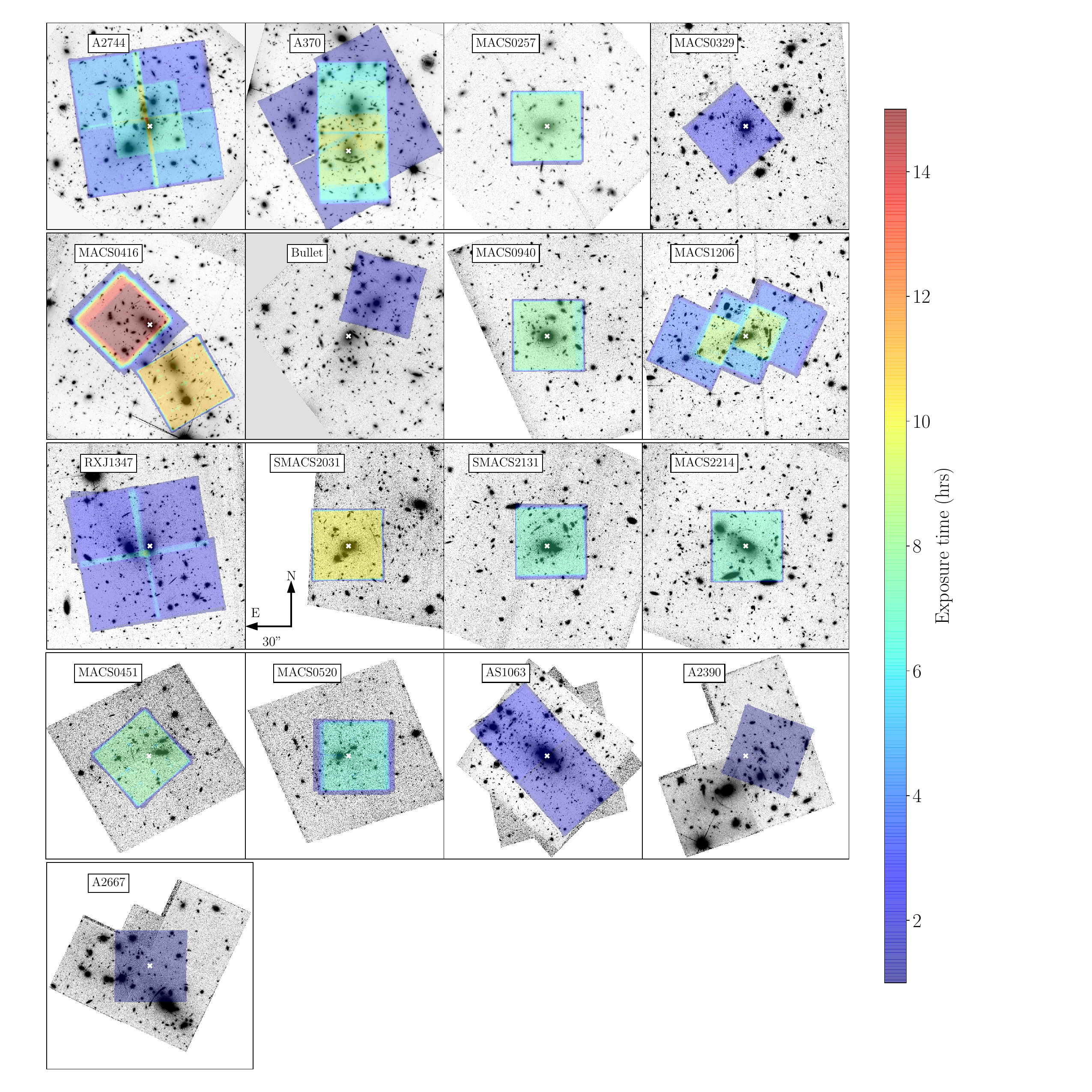}
   \caption{MUSE exposure maps overlaid by HST/F814W images adapted from \citetalias{Richard_2021}.}
              \label{exp_map}%
    \end{figure*}

The complementary HST data are, in addition for building lens models of clusters mentioned earlier, necessary to help with the detection of sources. The deep auxiliary HST data are listed in the Table \ref{table: cluster_info}. Seven clusters belong to the CLASH \citep{Postman_2012} and Frontier Fields \citep{Lotz_2017} programmes which are High Level Science Products (HLSP) combining all observations in 12 and 6 filters respectively. Absolute astrometry calibration is applied for these datasets with accuracy $<0".06$. The other clusters are part of the MACS survey and follow-up programmes (PIs: Bradac, Egami and Carton). They are calibrated using source catalogs from Gaia Data Release 2 \citep{Gaia}. Fig. \ref{exp_map} shows MUSE exposure maps of the 17 clusters overlaid on HST/F814W images adapted from \citetalias{Richard_2021}.
 
These data have been used for building spectroscopic catalogues providing prior source positions, which helps to distinguish overlapping sources detected in the MUSE cubes.

 \subsection{Identification of Lyman Alpha Emitters}

The construction of MUSE spectroscopic catalogues is documented in \citetalias{Richard_2021}. To detect line emissions, we make use of Muselet (MUSE Line Emission Tracker), a Python package written by J. Richard (\url{https://mpdaf.readthedocs.io/en/latest/muselet.html}). This in turn makes use of SExtractor \citep{Bertin} to detect emission line objects on MUSE Narrow Band continuum subtracted images. To account for low surface brightness sources such as some extended LAEs, a set of given SExtractor detection parameters (DETECT$_{-}$MINAREA, DETECT$_{-}$THRESH) have been used to make sure these faint sources are extracted properly. This set of parameters varies among these clusters. Sometimes a spatial filter is also used by Muselet for smoothing NB images. A spatial filter (5$\times$5 tophat with FWHM=0".8) has been used for A2390, A2667, MACS0257, MACS0451, MACS0520, MACS0940, MACS2214, SMACS2031, SMACS2131. Another one, (5 $\times$ 5 tophat with FWHM=0".4) has been used for A2744 and A370. 

 We also make use of the Source Inspector package, which is designed by the MUSE collaboration \citep{Bacon_2023}, using a combination of both deep HST and MUSE observations to identify sources and their redshifts. Depending on the nature of emission line objects such as signal-to-noise ratio (SNR) and clarity of line absorption/emission features, redshift confidence levels are assigned. A source with $zconf = 1$ is considered as a tentative case due to either a low SNR or the presence of ambiguous emission lines that do not display a typical Lyman alpha asymmetry profile or various low SNR absorption properties. A source with $zconf = 2$ is assigned based on emission lines having a higher
SNR compared to the previous case. Such sources may display several absorption lines with low SN. The confidence level will be upgraded if the source belongs to the multiple imaged systems. The highest level $zconf = 3$ is for the sources with a high SN of Lyman alpha line displaying a typical asymmetry shape and is accompanied by several emission/absorption lines showing its right typical profile. The catalog of sources of the 17 clusters including redshift and spectral line information can be accessed from \url{https://cral-perso.univ-lyon1.fr/labo/perso/johan.richard/MUSE_data_release/} (see \citetalias{Richard_2021}, for more detail). In the present work, there are such 190 unique LAEs categorized as $zconf = 1$. However, for our analysis we only consider LAEs with secure redshifts ($zconf =$ 2 and 3), excluding all tentative sources with a $zconf$ of 1. The identification of LAEs is described in detail in \citetalias{Vieuville_2019}, and is also mentioned in \cite{Claeyssens_2022} .

It is important to note, concerning sources detected both by HST, called P (prior) sources and by MUSE, M sources, that \citetalias{Richard_2021} kept the spatial positions of the P sources. We found an average position offset $\sim$ 0.2" between P and M sources. As the LAEs are detected by MUSE, we naturally use the M positions for these sources. 

Around 1350 Lyman alpha images with $zconf>=1$ have been identified behind the 17 lensing clusters. Many of them are multiple image systems. To avoid counting them more than once, for each multiple image system, we choose a representative image. This also helps to avoid de-lensing them which is computationally expensive. Choosing the representative for each multiple system is manually performed by investigating them one by one. The HST images (\url{https://archive.stsci.edu/missions/hlsp/clash/} and \url{https://archive.stsci.edu/missions/hlsp/frontier/}) and MUSE White Light Image are useful for this selection process. If possible, the chosen representative has high enough S/N, reasonable magnification errors, and a position which is well isolated, with little to no contamination from nearby bright sources. Finally, 600 LAEs sources have been selected at this stage with, in some cases, several different images equally good to represent the parent source. All these sources have $zconf=2$ or $3$. Their redshift distribution is shown in Fig. \ref{fig: combine_3images} together with those from previous work of \citetalias{Vieuville_2019} for comparison. A large number of LAEs in the present work have redshifts in the range of 2.9 - 4.0. There are two bumps in the distribution at redshifts of $\sim$4 and $\sim$5, suggesting some overdensities along the line of sight.

\section{Modeling lensing clusters mass distributions}\label{sec: 3}
\subsection{Mass distributions by LENSTOOL}
There are two approaches in lens modelling: parametric mass modelling and free-form (non-parametric) methods. For the present work, we use the former to build the total mass distribution of lensing clusters, based on Lenstool \citep[see e.g.][]{Kneib_1996,Jullo_2007, Jullo_2009}. Following Lenstool, and the procedure described in \citetalias{Vieuville_2019} and \citetalias{Richard_2021}, there are two components that contribute to the lensing. One is the large scale structure of the cluster, the other is from each massive cluster member. The contribution and implementation of each component is well described in \cite{Jullo_2007}. 

Parameters of individual mass profiles are sampled via Monte Carlo Markov Chain (MCMC) to compare the match between the observed and predicted multiple images. Details of lens modeling are described in  \citetalias{Richard_2021}. The lens models and their parameters are listed in Table \ref{table: lens model}. Twelve out of seventeen lens models used in the present work come from the work of \citetalias{Richard_2021}. Thanks to these models, one can work back and forth between the image and source planes for all LAEs. This allows us to correct for the amplification factor, i.e. lensing magnification, for each source, as well as the source plane projected area allowing us to capture the extended morphology of LAEs.
\begin{figure}[!h]
   \centering
   \includegraphics[height=9.cm]{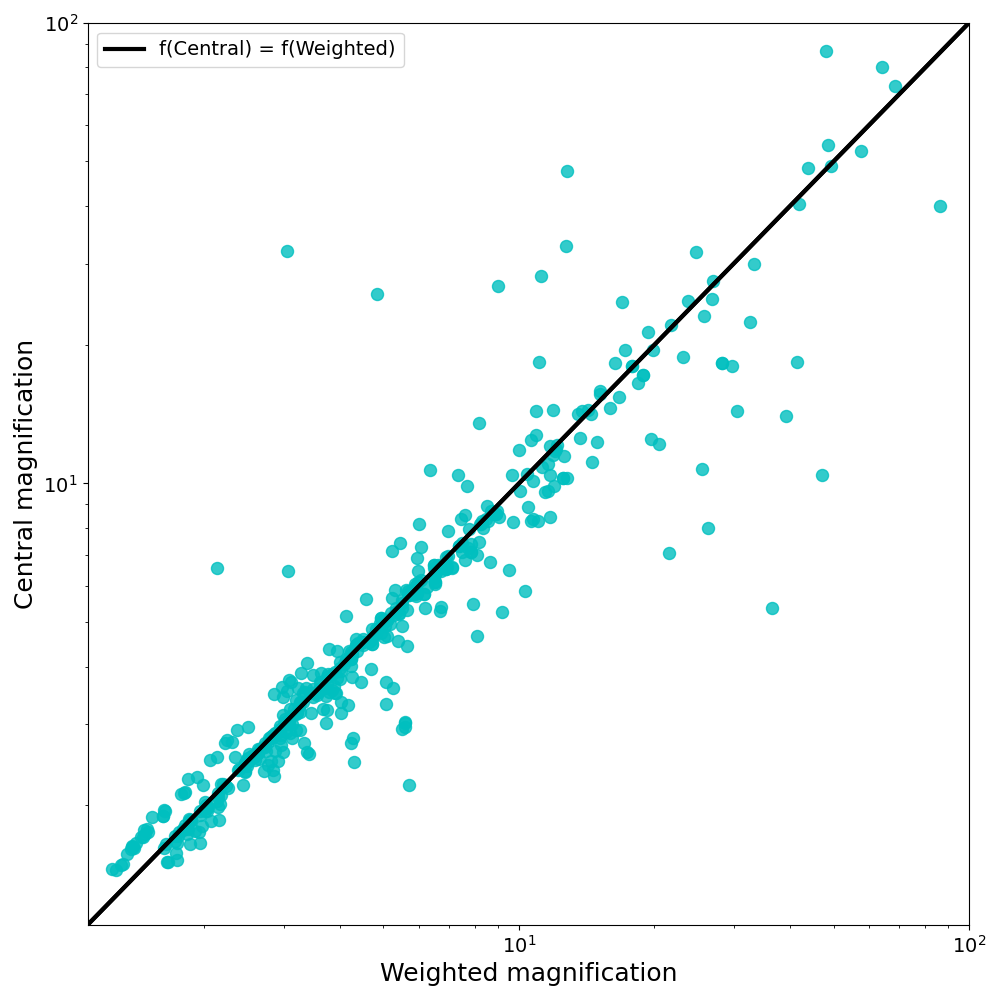}
   \caption{Source central magnification vs. its flux-weighted average one for LAEs in the sample. Extended sources often have averaged magnification values lower than those of their central ones, as expected. The black dashed line denotes a one to one magnification ratio.}
              \label{magnification_compare}%
    \end{figure}
Note that magnification associated with a given source obtained from Lenstool is the magnification of the source centre, not the entire source. If the LAE in question is a compact one, and is relatively far from the lens caustic lines, then it is fair to assign this magnification for that source. Such is the case for 90$\%$ sources in the sample. However, for extended sources, in particular those close to the lens caustic lines, magnification strongly varies across the source and using the central point magnification is not justified. In such cases, one must take the flux-weighted magnification by averaging the magnification over the entire source. In this work, we use this weighted magnification for all LAEs. We can do so systematically by using the segmentation map which is created from the NB image by Sextractor and is used as a mask for NB fluxes of the LAEs, to switch back and forth between the image and source planes. Fig. \ref{magnification_compare} displays the difference in magnification between the central and the flux-weighted values for the LAEs in the present sample. 

\begin{figure*}
   \centering
   \includegraphics[height=12.5cm]{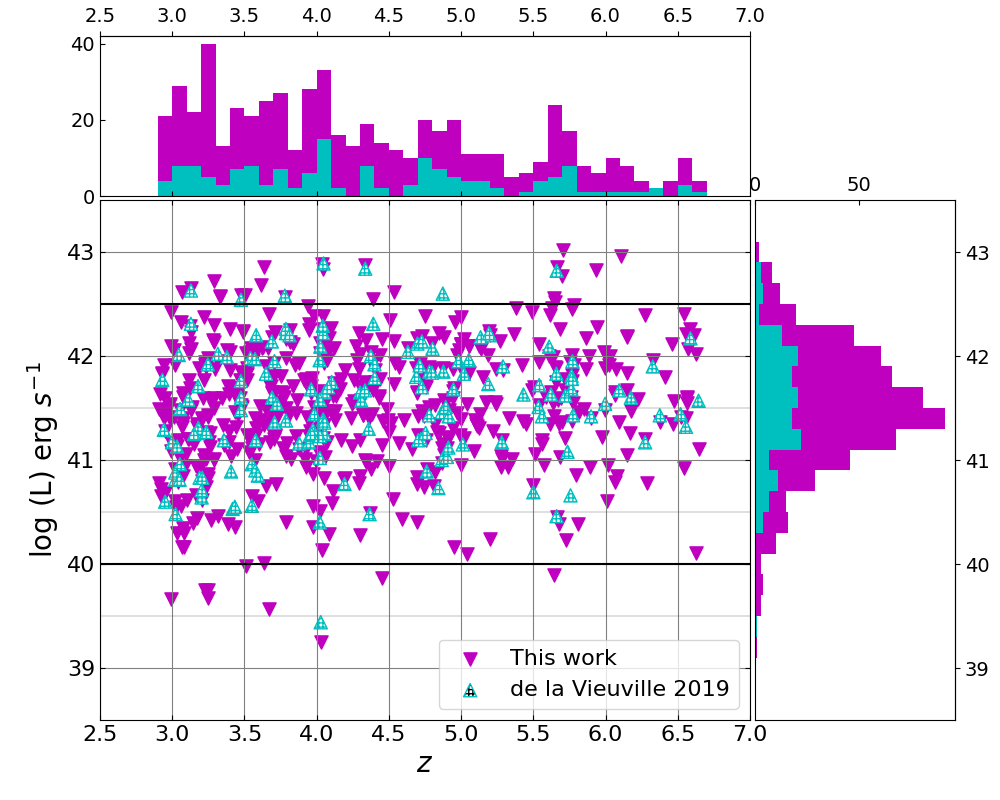}
   \caption{Lyman $\alpha$ luminosity versus redshift of lensed LAEs in the present work (magenta) and those of
   \citetalias{Vieuville_2019} (cyan). Histograms of redshift (top) and of luminosity (right) are also shown, respectively.}
             \label{fig: combine_3images}%
    \end{figure*}

\subsection{Flux measurement}
  
We have considered two methods to measure fluxes of LAEs. Fluxes are either extracted by fitting the Lyman$-$alpha profile as performed by \cite{Claeyssens_2022} or by running SExtractor with FLUX$_{-}$AUTO on the NB images in which the LAEs are detected as performed by \citetalias{Vieuville_2019}. The first method is employed for LAEs behind 16 lensing clusters except for the A2744 while the second method is applied for the cases in which the first method fails to fit the Lyman alpha profile. Using the first method, we obtained flux measurements for 425 LAEs. However, 47 faint source fluxes could not be measured using this approach. In such cases, we employed the SExtractor package (second method) to extract the fluxes. A large fraction of these failed sources has a low SNR but they have been upgraded to $zconf=2$ because they are members of multiple imaged systems. Furthermore, we used the second method to extract LAE fluxes for sources located behind the lensing cluster A2744. The reason for this is that the source offset positions of the M sources are not available in the new catalogs provided by \cite{Claeyssens_2022} whereas the results in \citepalias{Vieuville_2019} are in line with the other literature. Overall, we obtained 425 source fluxes using the first method and 175 source fluxes using the second method. We expand briefly on both methods below.

Regarding the first method,  \cite{Claeyssens_2022} extract LAE fluxes via three main steps: spectral fitting, NB image construction and repeating spectral extraction. Firstly, a pseudo NB image of each slice (measuring 5" $\times$ 5") is constructed from the MUSE datacube. They use the formulae introduced by \cite{Shibuya_2014} to fit the asymmetric spectral profile of LAEs: 
 \begin{equation}
 \label{func: shibuya}
     f(x)=A \text{ exp}\left[\frac{-(\lambda -\lambda_0)^2}{2(a_{asym}(\lambda-\lambda_0)+d)^2}\right]
 \end{equation}
where $A$ is the flux amplitude, $\lambda_0$ is the wavelength layer in which source's emission reaches  maximum, $a_{asym}$ accounts for the asymmetry of the Gaussian function, and $d$ is the width of the line. The mean value of $d$ is $7\,\AA$, and the mean value of $a_{asym}$ is $0.20$. The value of $\lambda_0$ is converted directly from source redshift and the flux amplitude $A$ is integrated in the range [$\lambda_{0}-6.25$: $\lambda_{0}+6.25$]$\,\AA$. 

It is worth mentioning that 31 out of the 600 LAE spectra display similar 2-peak profiles. We apply the same procedure as described above, dealing with the blue and red peaks separately, taking into account the amplitude difference between the two. The two peaks are then summed to get the final flux. The flux uncertainty is estimated using python package EMCEE \citep{Foreman-Mackey_2013} with 8 walkers and 10,000 iterations. 
   \begin{figure}[!h]
   \centering
   \includegraphics[height=8cm]{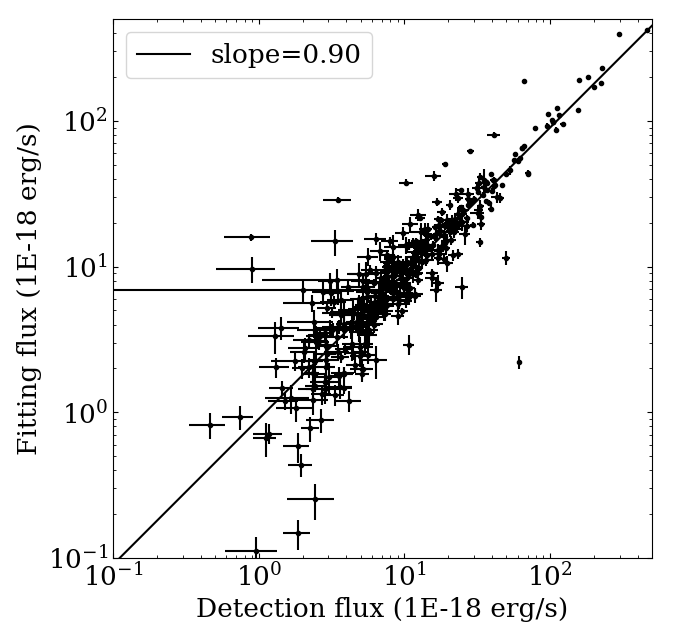}
   \caption{Fluxes of LAEs in the sample, obtained by fitting Lyman-alpha profiles \citep{Claeyssens_2022} versus those extracted by SExtractor. Solid black line indicates the best linear fit between two the fluxes. }
   
              \label{fig: flux_compare}%
    \end{figure}
For the second method, we make use of the SExtractor software \citep{Bertin} running on NB detection images which host LAEs. This is well described in \citetalias{Vieuville_2019}. For this purpose, a sub-cube with a size of 10" $\times$ 10" $\times$ the width of the Lyman-$\alpha$ emission is first extracted from the original cube, then averaged to form the LAE image. Two sub-cubes of the same spatial size, extending 20 $\AA$ bluewards and redwards of the Lyman-$\alpha$ line are also created. These are then averaged to form the mean continuum image of each LAE. By subtracting this mean image from the LAE image, pixel by pixel, one forms a new measurement image which is ready for SExtractor. We use SExtractor's FLUX$_{-}$AUTO parameter to estimate the LAE fluxes together with their uncertainties. This parameter is developed based on Kron's first moment algorithm: when a given source is convolved with a Gaussian seeing, 90$\%$ of its flux will be inside the circular aperture of the Kron radius. This holds even for extended sources. The final flux obtained from the measurement image is then multiplied by the width of the Lyman-alpha emission. 

It may happen that some sources are either faint or have low extended surface brightness. In such cases, SExtractor does not extract these sources properly. A set of farsighted SExtractor parameters (DETECT$_{-}$THRESH, DETECT$_{-}$MINAREA), have been proven to be very useful to overcome this problem. SExtractor progressively releases detection conditions using this set of parameters to make sure that faint sources can be extracted. 

Fig. \ref{fig: flux_compare} shows a comparison between the fluxes of LAEs obtained by using the two methods in the present work. In general, the two fluxes agree well with each other within a range of a few orders of magnitudes. The deviation spreads out at the faint fluxes, as expected, but no systematic trend is observed. The source fluxes, obtained from the second method, will be used later to estimate the systematic uncertainties attached to the faint end slope.

As mentioned in the previous subsection, to obtain the intrinsic luminosity of a source, its segmentation map is projected into the source plane using Lenstool, then the weighted (averaged) magnification over that source is calculated. The weighted magnifications obtained this way are used throughout this work.
The source intrinsic luminosity is computed as:\\
\begin{equation}
     L_{Lya}=\frac{F_{Lya}}{\mu} 4\pi D^2_L
   \end{equation}
where $F_{Lya}$ is the flux of the LAE obtained as described above, $D_L$ is luminosity distance and $\mu$ is weighted average magnification. The luminosity versus redshift and luminosity histogram  of our sample together with the ones from \citetalias{Vieuville_2019} are shown in Fig. \ref{fig: combine_3images}.

\begin{figure*}[!htb]
   \centering
   \includegraphics[height=6.5cm]{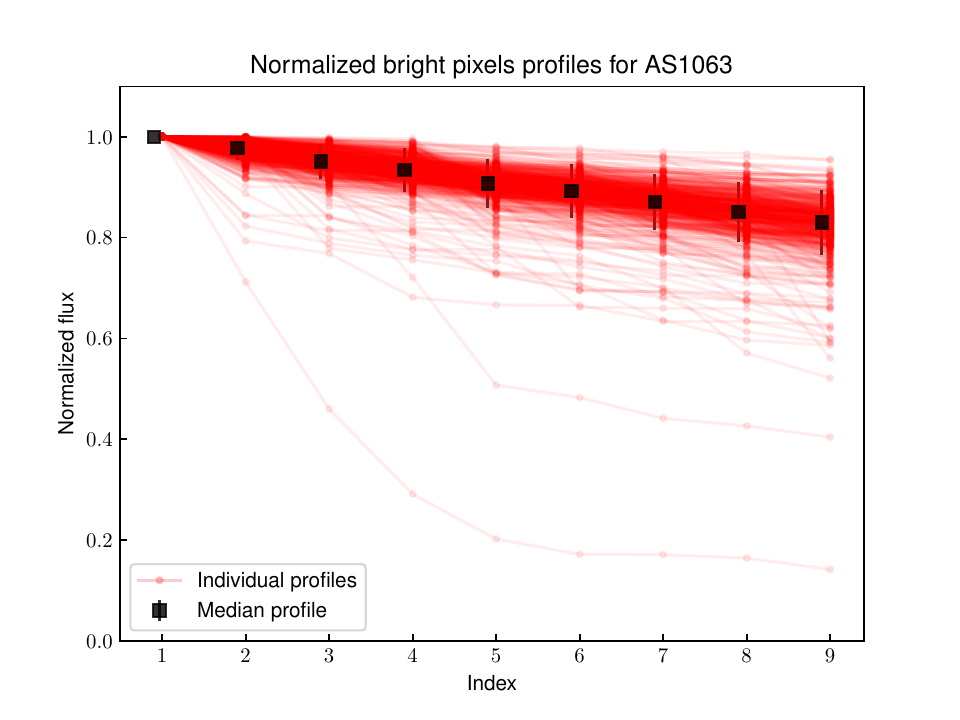}
   \hspace{-0.cm}\includegraphics[height=6.5cm]{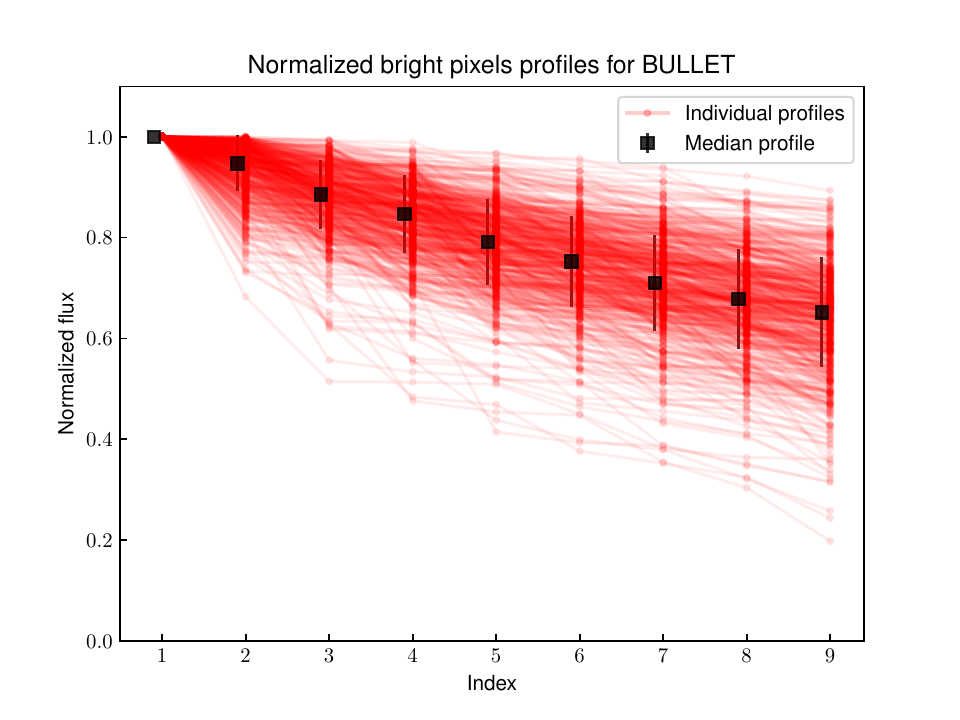}
   \caption{Examples of normalized bright pixels profiles of sources from clusters AS1063 and BULLET. Individual profiles (red) are shown together with their median (black). Each of these median profiles is then used as a representative for all sources in that MUSE datacube. The different seeings between datacubes are taken into account when producing these profiles. The median profile and the RMS map are used to produce a wide range of S/N for each MUSE datacube.}
              \label{Bpg_profile}%
\end{figure*}

\section{Building the LF}\label{sec: 4}
\subsection{The $V_\text{max}$ method}
In order to deal with the complexity of the lensing datacubes obtained by MUSE, we adopt a non-parametric approach, the $1/V_\text{max}$ method \citep{Schmidt_1968}, allowing us to treat each source individually, to build the LF using detected LAEs behind lensing clusters. The $V_\text{max}$ of a source is the volume of the survey where that individual source could be detected. Inverting this gives the contribution of that source to a number density of galaxies. Determining $V_\text{max}$ is the crucial step in building the LF and this is challenging. Noise is an important factor. The $V_\text{max}$ has to take into account noise introduced by the instrument, by sky line variations for different wavelength layers and by local noise variations for different sky directions in the MUSE FOVs. It must take into account the source detectability in all the MUSE lensing datacubes involved in the survey, i.e. the whole survey volume. Additionally, it has to be calculated in the source plane to avoid adding the same survey volume for multiple image systems. All of these steps make $V_\text{max}$ determination complex and computationally expensive.

\citetalias{Vieuville_2019} developed a package to compute $V_\text{max}$ of lensed LAEs behind lensing clusters observed by MUSE. Details of the algorithm are described in \citetalias{Vieuville_2019}. Here we briefly summarize the important points and main steps of the procedure.

Detectability masks are an important concept for computing $V_\text{max}$. They are built first in the image plane, by comparing a given source brightness distribution pixel by pixel to the RMS maps, channel by channel, in the spectral range of all 18 datacubes. These masks are projected in the source plane using Lenstool, to obtain source plane masks. However, performing this for all sources is extremely computationally expensive. The main important point proposed by \citetalias{Vieuville_2019} is to build a set of pre-computed 2D detection masks that cover a wide range of signal to noise ratios for each MUSE datacube. Hence, for a given source, one first computes its S/N in the parent cube, in which the source is detected, then picks up the corresponding 2D mask from the pre-computed list available, which is closest match with the input in terms of S/N. One can do so given that for each lensing datacube root mean square (RMS) maps of different layers display roughly the same type of patterns. One can choose a representative of these RMS maps, i.e. their median map, then scale it accordingly by a constant factor to get other RMS maps for different layers. Another important simplification is that, for each lensing cluster, one uses a representative brightness profile for all the sources, i.e., the median of radial source brightness profiles. Using both median radial profile and median RMS map, one then produces masks at wide range of S/N covering the source sample. This is another simplification to limit the number of masks created in both the image and source planes. 
\begin{table}[!h]
\centering
     \caption[]{Total co-volume of 17 clusters $(2.9<z<6.7)$}
        \label{tab: volume_info}
        \begin{tabular}{lcc}
            \hline
            \hline
            \noalign{\smallskip}
       Cluster &Total co-volume [Mpc$^3$] \\
        \noalign{\smallskip}
        \hline
        A2390  &735\\
        A2667  &885\\
        A2744  &  10500 \\
        A370   &5350\\
        AS1063   &1970\\
        BULLET   &895\\
        MACS0257 &730\\
        MACS0329 & 1225\\
        MACS0416N   & 3420\\
        MACS0416S   & 1670\\
        MACS0451    &1210\\
        MACS0520    & 765\\
        MACS0940& 5760 \\
        MACS1206  &2980\\
        MACS2214    & 1100\\
        RXJ1347&  7920\\
        SMACS2031   & 1675\\
        SMACS2131    &  920   \\
        \hline
        \textbf{Total:}  & \textbf{49710}\\
           \noalign{\smallskip}
            \hline
            
        \end{tabular}
     
  \end{table}
  
The Lenstool package is used to perform the reconstruction from image plane to source plane. To account for variations of projected source plane area as a function of redshift, each 2D mask reconstruction is sampled at four different redshifts, $z$ =3.5, 4.5, 5.5 and 6.5. One of the four source plane masks is then assigned for a given source, i.e. the one is the closest match with its redshift. Source plane magnification maps are also used to eliminate regions with insufficient amplification implying sources could have not been detected there. 3D masks for a given source are formed in this way. The ﬁnal volume is integrated from the unmasked pixels of these 3D source plane masks:
\begin{equation}
    \mathrm{V}=N \omega \frac{c}{H_o} \int_{zmin}^{zmax} \frac{D^2_L(z)}{(1+z)^2E(z)} dz
\end{equation}
where $N$ is number of unmasked pixels counted on all wavelength layers, $\omega$ is the angular size of a pixel, $D_L$ is the luminosity distance, and $E(z)=\sqrt{\Omega_m (1+z)^3+(1-\Omega_m -\Omega_{\Lambda})(1+z)^2+\Omega_{\Lambda}}$.

\subsection{Completeness value}
Completeness is a correction related to the chance that a given source is detected at its own wavelength affected by random variation of noise on the spatial dimension of the channel i.e. the NB layer where the source's emission reaches maximum. Together with $V_{\text{max}}$, it is a crucial correction before deriving LF points. Depending on source morphology, this completeness is computed individually. Details on how to compute completeness are described in \citetalias{Vieuville_2019}. We condense this, as follows:

- A source profile of individual LAE is created from a combination of filtered image, object image, and segmentation image obtained by running SExtractor on the peak NB layer. 

- To compute the completeness for a given source, one has to make simulations by randomly injecting the source on the masked NB image layer where the emission reaches maximum. We use the real source profile to perform this task. The local noise in the region where the source has been injected is likely to decide whether or not a source is detected. SExtractor scans on the masked NB image, with exactly the same parameters as when that source was originally extracted. We iterate 500 times for each source, and the completeness is the detection success rate.

- The quality of extraction by SExtractor is also taken into account. Following \citetalias{Vieuville_2019}, sources with flags of type 1 and 2 are trustworthy, but sources with flag 3 are doubtful and will not be used for the LF computation afterwards. From the original sample of 600 LAEs, 588 with flags 1 and 2  are kept at this stage (see below).

\begin{figure}[!h]
   \centering
   \includegraphics[height=6.5cm]{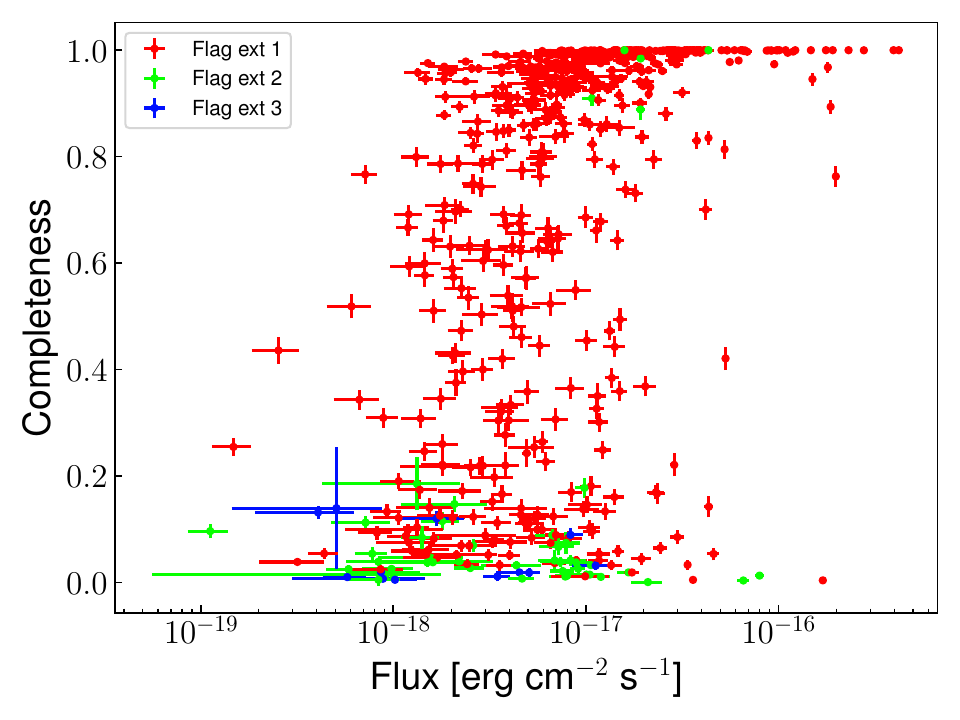}
   \caption{Completeness vs. detection flux of LAEs from the present sample. Different colors indicate the quality of extraction from SExtractor. Only sources in the unmasked regions of the detection layer are considered here. }
              \label{fig: comp_map}%
    \end{figure}

\begin{figure}[!h]
  \centering
   \includegraphics[height=6.5cm]{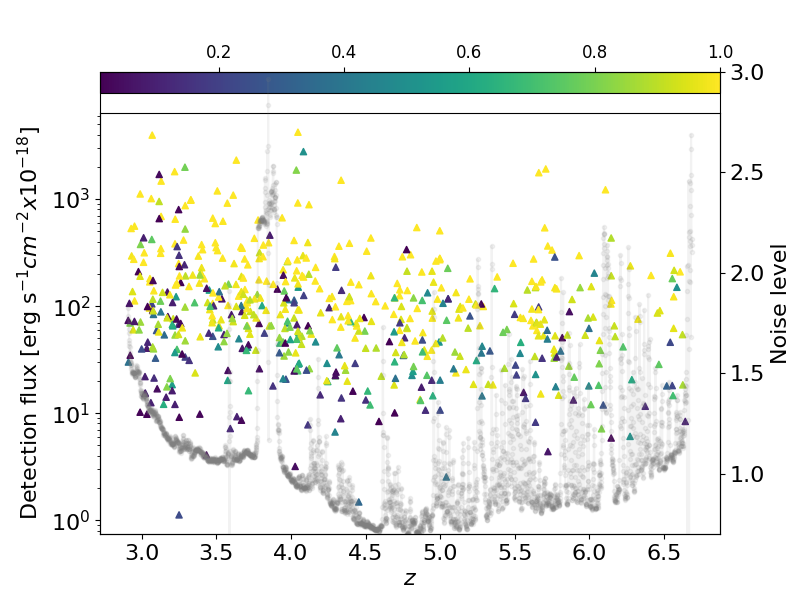}
  \caption{Detection flux vs. redshift for all sources in the present sample.  The color bar on the top shows completeness value of each source in the sample. The grey points show the evolution of noise level as function of wavelength/redshift for the cluster MACS0257 as an example.}
              \label{Fig: correlation flux redshift completeness normal}%
\end{figure}

\begin{figure}
   \centering
   \includegraphics[width=0.5\textwidth]{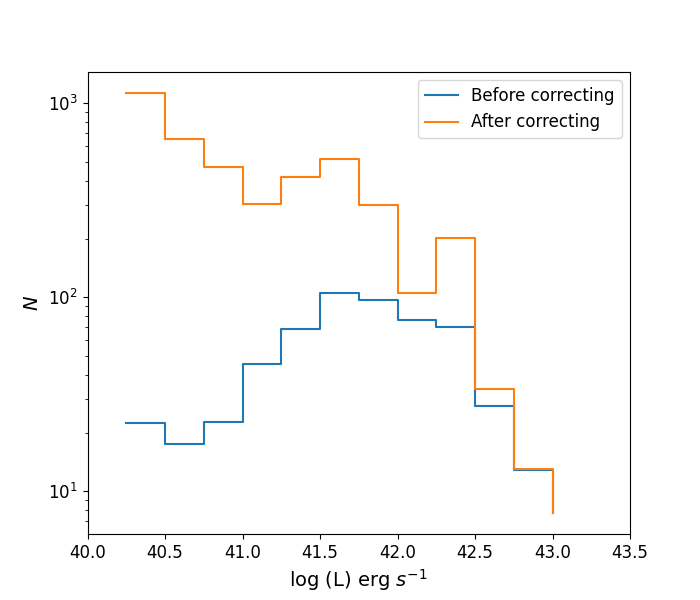}
   \caption{
      Source distribution in each luminosity bin (the faintest on the left, the brightest on the right) before and after correcting for the completeness. A larger correction, a factor up to a few tens, has been made at the faint luminosity regime, as expected.}
              \label{Fig: Source_distribution}%
    \end{figure}       
    
We found that the size of the masked NB images used to re-detect mock sources has an impact on the quality of extractions. When a source has a few neighbors, the local noise is not well represented for that wavelength slice if the size of the image is not large enough. To deal with this, we have increased the size from 30"$\times$30" (as in \citetalias{Vieuville_2019}) to 80"$\times$80". This is more computationally expensive but better accounts for the local noise. The extraction results for the sample are: 555 cases with flag 1, 33 with flag 2 and 12 with flag 3. Additionally, aiming to include as many sources as possible for calculating the contribution to cosmic re-ionization, all sources with completeness values above 1$\%$ are kept. Taking this into account, 575 LAEs are used for the next steps.

Fig. \ref{fig: comp_map} shows the completeness of each LAE vs. their respective detection flux in the sample. The average completeness value is 0.69 and the median value is 0.90 over the entire sample. The equivalent values from \citetalias{Vieuville_2019} are 0.74 and 0.9 respectively. Our average value is a bit smaller than that of the previous work due to numerous faint sources in our sample, which often have low completeness values.

For a given source, one may expect that higher its detection flux leads to higher completeness value. However, it is not always the case as shown in Fig. \ref{Fig: correlation flux redshift completeness normal}. Some sources still have low completeness values, i.e. below 0.2, while their detection fluxes are relatively high. It suggests that source morphology also plays an important role in its completeness, as previously mentioned by \citetalias{Vieuville_2019}. 

\subsection{Luminosity Function determination}

We divide the  575 LAEs into four redshift bins in order to study the evolution of the LF with redshift: $2.9<z<6.7$, $2.9<z<4.0$, $4.0<z<5.0$, $5.0<z<6.7$. The contribution of each individual LAE in each bin is calculated as follows:
\begin{equation}
    \phi(L_i)=\frac{1}{\Delta \text{log } L_i} \sum_j \frac{1}{C_j V_\text{max,j}}
\end{equation}
where $i$ corresponds to the luminosity bin, $\Delta \text{log } L_i$ is the width of the luminosity bin in log scale, $j$ corresponds to the source in the sample and $C_j$ and $V_\text{max,j}$ are the completeness value and survey volume of the source in bin $i$.

We use MCMC to calculate the mean and the statistical uncertainty for each LF point by randomly generating a set of magnification, flux and completeness values for each individual source. 20,000 catalogs were built from the original catalog. For each LAE, the flux and completeness are randomly sorted following Gaussian distributions having means as their measured values and sigmas as their uncertainties. A random magnification value is taken from the magnification distribution $P(\mu)$ of each source, accounting for extended source morphology and uncertainties in the lens model. With respect to \citetalias{Vieuville_2019}, we have introduced an improved procedure to determine the $V_\text{max}$ associated to a random value of the magnification. This allows us to better capture the variations in $V_\text{max}$ taking place for very large values of the magnification. For each iteration, a single value of the LF is obtained for each luminosity bin. The distribution of LF values in each bin, obtained in this way, is used at the end of the process to determine the median of $\phi (L_i)$ in each luminosity bin in linear space, and to compute the asymmetric error bars.  

For the estimation of the cosmic variance, we used the cosmic variance calculator presented in \cite{Trenti_2008}. A single compact geometry made of
the union of the effective (lensing-corrected) areas of the 18 FoVs is assumed following \citetalias{Vieuville_2019}.

\subsection{Schechter function fitting}\label{sec: Schechter fit}
After obtaining the LF points following the procedure explained above, we fit the results using a Schecter function. The Schechter funcion, proposed by \cite{Schechter_1976}, has been extensively used to describe the LF as well as its evolution with redshift. It is often written as:
\begin{equation}
    \Phi(L)dL =\frac{\Phi*}{L*}\left(\frac{L}{L*}\right)^\alpha \text{exp} \left(\frac{-L}{L*}\right) dL
\end{equation}
where: $\Phi*$ is a normalization parameter, $L*$ is the luminosity at the point where the power law changes to an exponential law at high luminosity, $L$ is the luminosity, $\alpha$ is the faint end slope and $\Phi$ is the number density in a given of logarithmic luminosity interval. 
  
Regarding the fitting method, we adopt the same procedure as described in \citetalias{Vieuville_2019}, using a Schechter function with three free parameters to vary ($\Phi^*$, $L*$, $\alpha$). We first minimize these parameters by using the Levenberg-Marquardt algorithm, in particular using traditional chi-square minimization procedure, provided by the standard package {\it Lmfit}. 

    \begin{figure}[!h]
   \centering
   \hspace{-0.6 cm}
      \includegraphics[height=5.7cm,width=8cm]{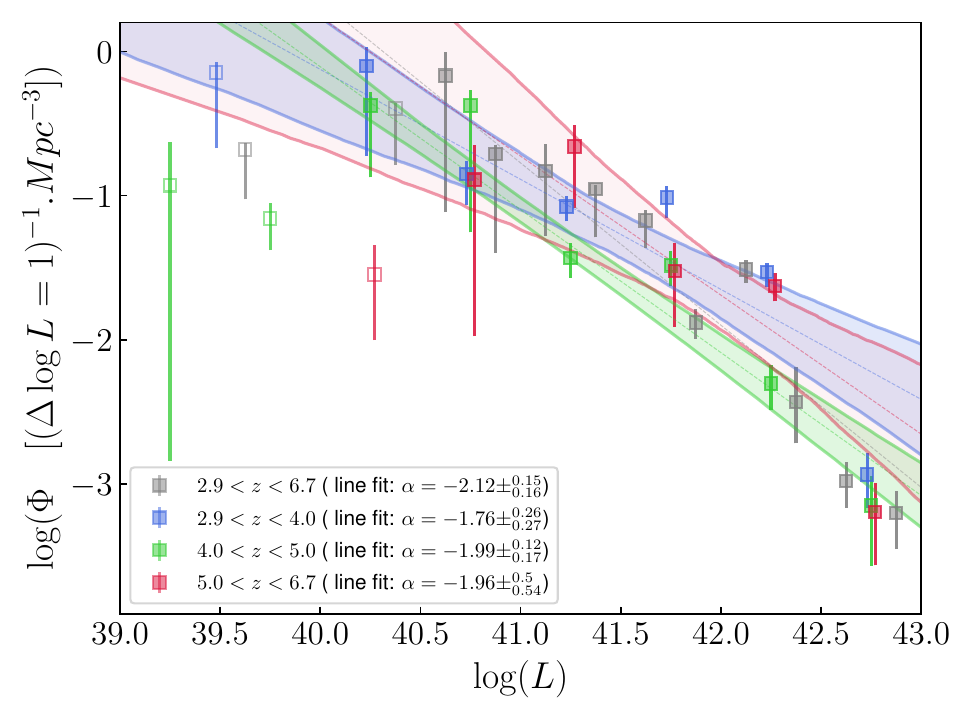}
   \caption{Faint end slope of LF: Line (power law) fitting for four different redshift ranges: 2.9 < $z$ < 4.0 (blue), 4.0 < $z$ < 5.0 (green), 5.0 < $z$ < 6.7 (red), 2.9 < $z$ < 6.7 (black). These fits use only our LF points constructed from the current sample. Open symbols at the faintest luminosity bins are not included in the fitting process.}
              \label{fig: LF in 68 percent}%
    \end{figure}
   
\begin{figure*}
   \centering
    \includegraphics[height=6.8cm]{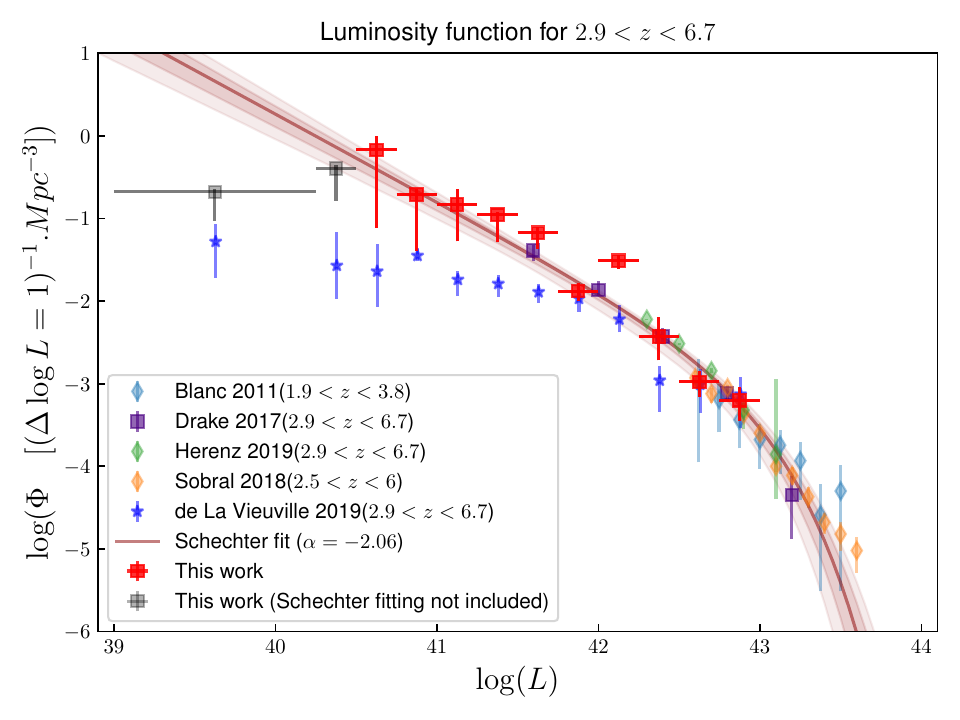}
    \includegraphics[height=6.8cm]{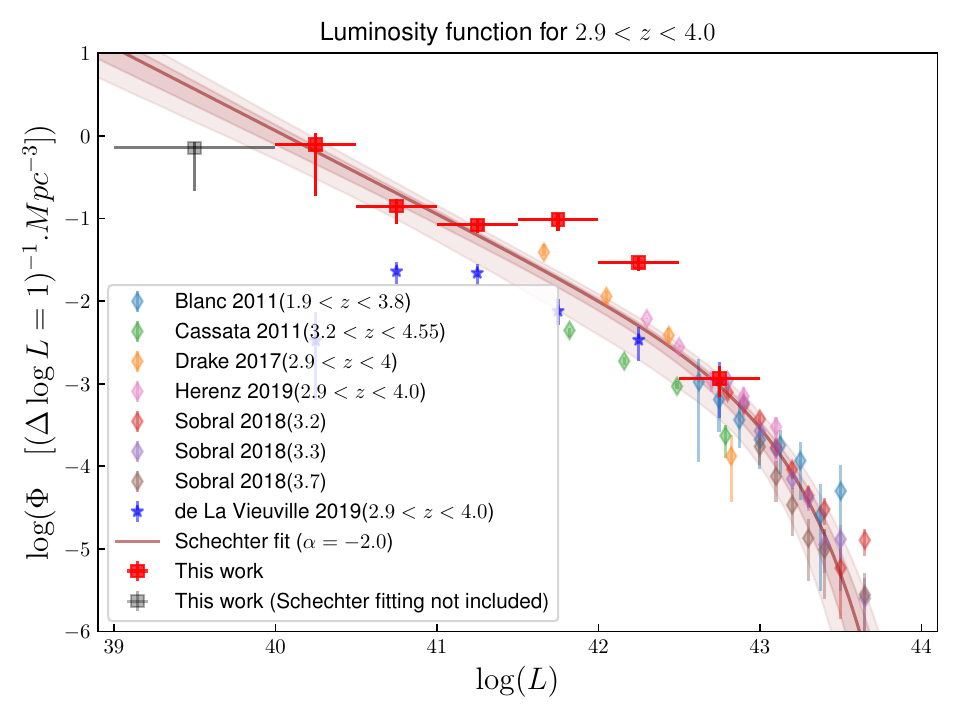}
    \includegraphics[height=6.8cm]{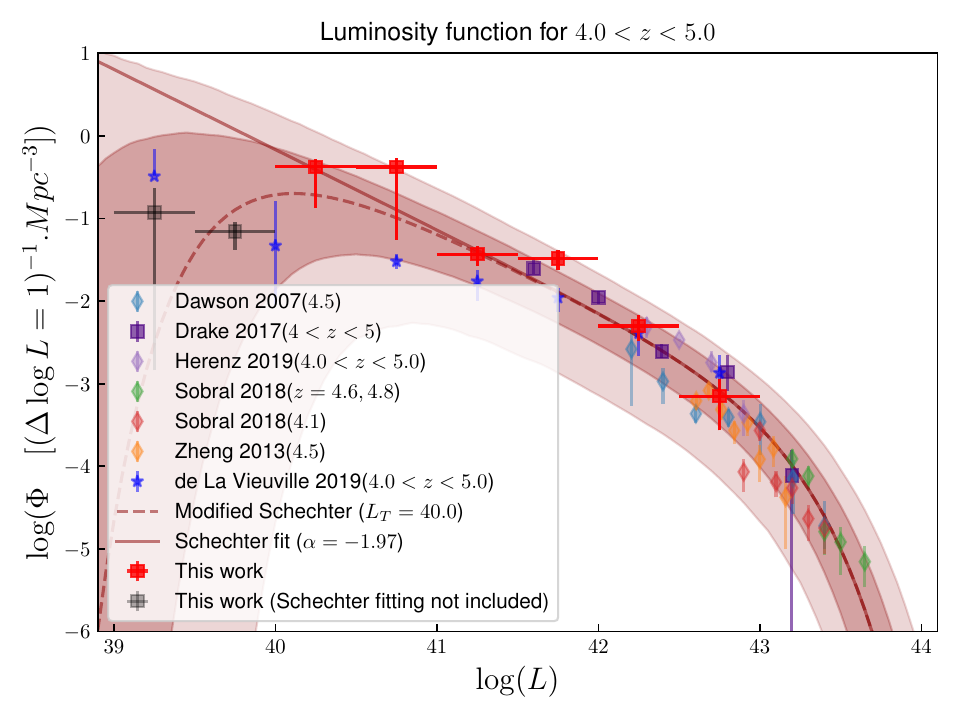}
    \includegraphics[height=6.8cm]{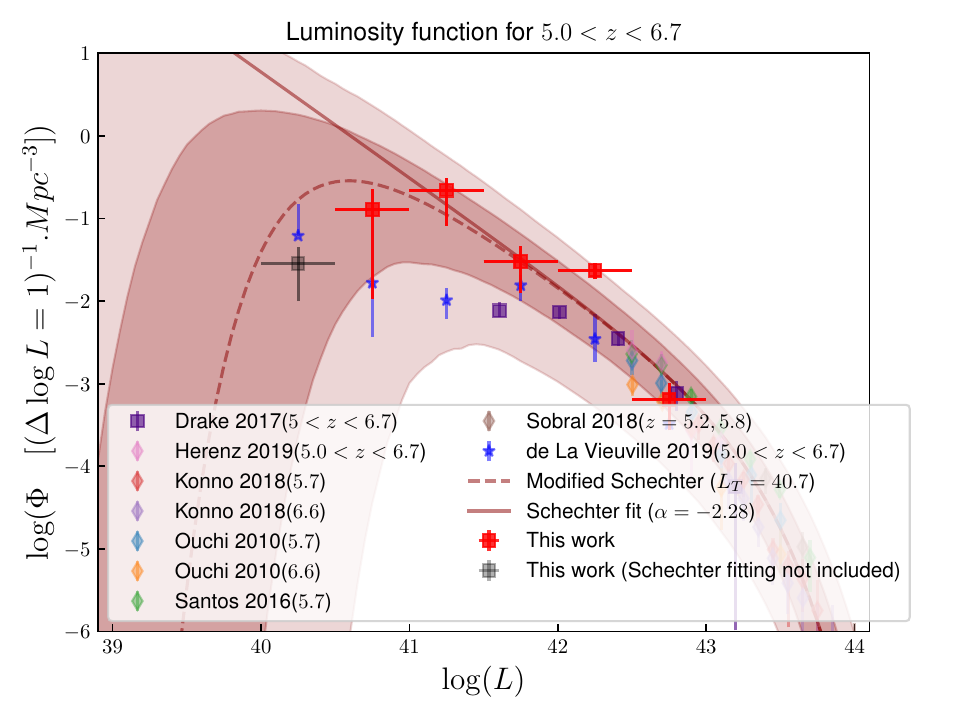}
    \caption{LF points and their fits for different redshift intervals including previous literature data points. The red squares are the points from the present work. The literature points at the bright end of the LF: \cite{Ouchi_2010}, \cite{Sobral_2018}, \cite{Zheng_2013}, 
   \cite{Herenz_2019} and 
   \cite{Drake_2017} have been used for the fitting. The blue points \citetalias{Vieuville_2019} are shown for comparison purpose only. The best fits (Schechter function) are shown as a solid line and the 68\% and 95\% confidence areas as dark red coloured regions, respectively. The dashed lines shown in lower panels are modified Schechter functions to account for a possible turnover at faint luminosity bins (see text).}
              \label{fig: LF in different redshift bin}%
    \end{figure*}

    \begin{figure*}
   \centering
   \centering
   \hspace{1cm }\includegraphics[width=0.45\textwidth]{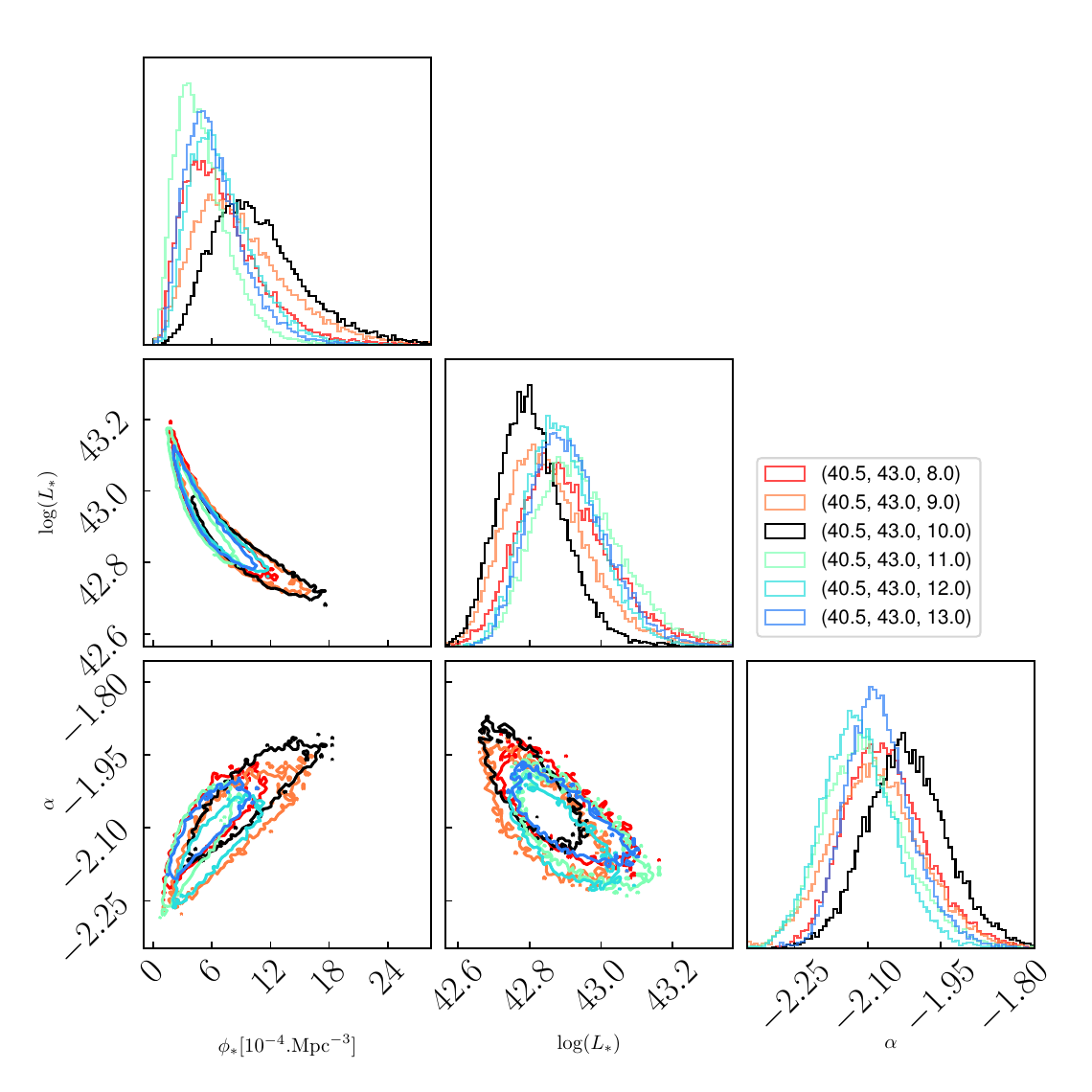}
    \hspace{0.5cm}\includegraphics[width=0.45\textwidth]{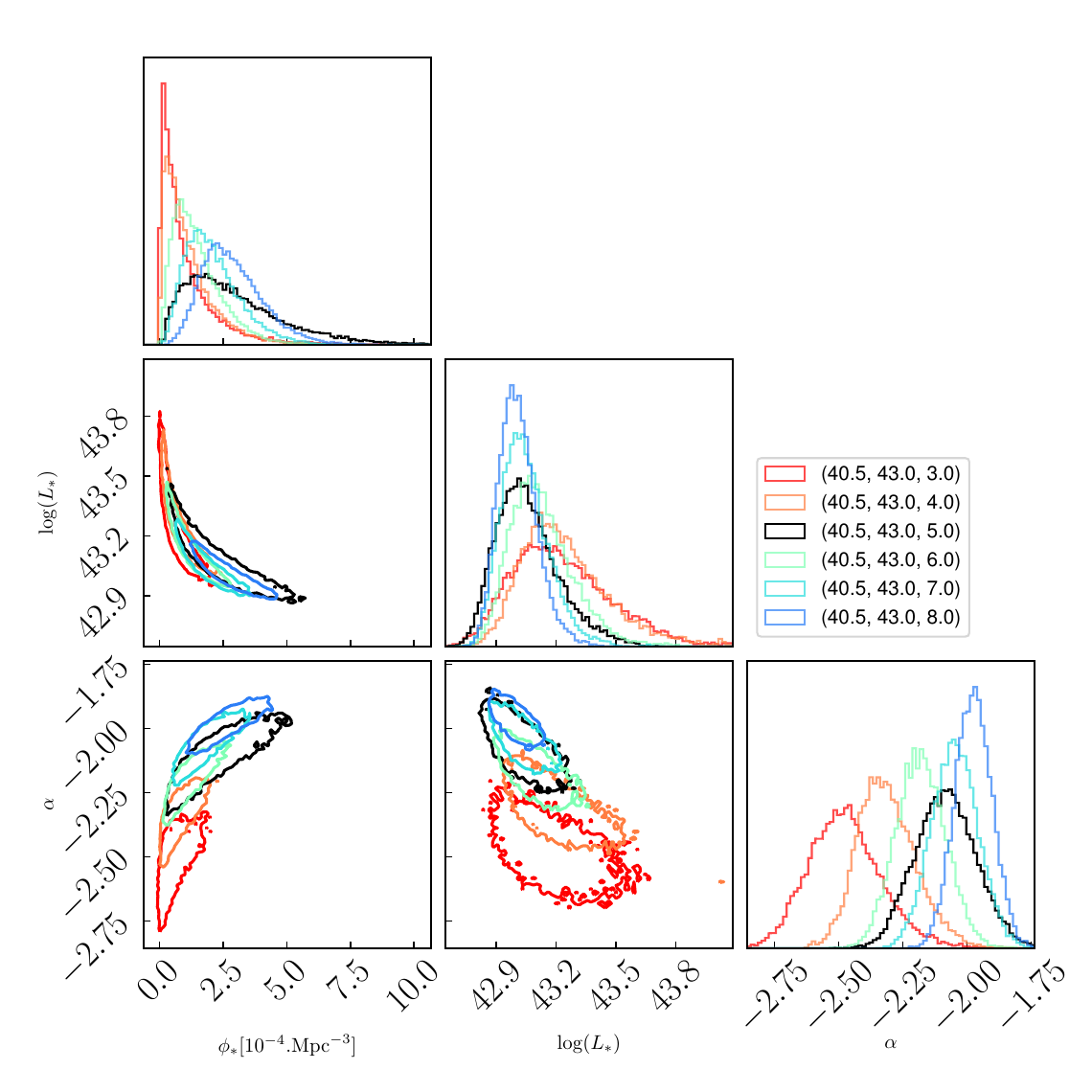}
   \caption{Correlations between three Schechter's best-fit parameters for different luminosity bins. The tuples denote the lower and upper limit of the luminosity range with respect to the number of the bin. Contours are 68\% confidence levels obtained from the Schechter's fitting. Results are shown for 2.9 < $z$ < 4.0 (left) and 5.0 < $z$ < 6.7 (right).}
              \label{fig: LF in different luminosity bin}%
    \end{figure*}
\begin{figure}
   \centering
   \centering
   \hspace{0.1cm}
   \includegraphics[width=0.5\textwidth]{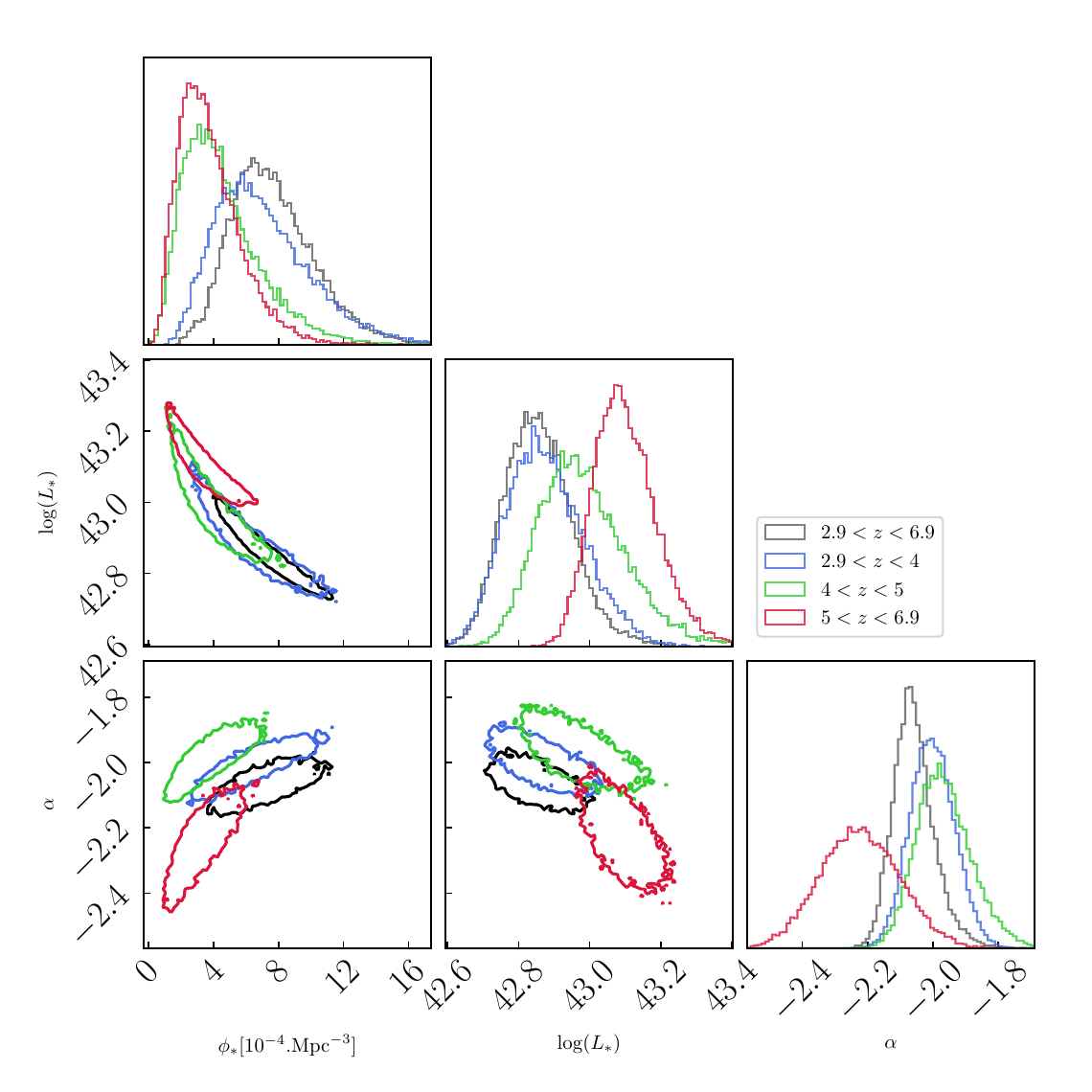}
   
   \caption{Correlations between the three  parameters, $L*$, $\Phi*$ and $\alpha$ of the Schechter function for four different redshift bins as indicated in the inserted. Contours show 68$\%$ confidence levels obtained from fitting procedures.}
              \label{fig: 3 parameters correlation}%
    \end{figure}

 \section{The faint end of the LF}\label{sec: 5}
 
 \subsection{Faint end LF}\label{sec: 5.1}

The faint-end slope of the LF function is still a matter of debate and more observations are strongly needed.
Strong gravitational lensing by clusters of galaxies is of great help to probe the faint galaxy luminosity regime and hence to constrain the shape of LF in that region. The price to pay is that the cosmic volume of the survey is significantly reduced: the higher the lensing magnification factors, the lower the volume probed. Our sample efficiently probes down to 10$^{40}$ erg s$^{-1}$, the same faint luminosity regime as studied by \citetalias{Vieuville_2019} but with improved statistics and a better coverage at the faint end of the luminosity distribution (see Fig. \ref{fig: combine_3images}) and 34'000 Mpc$^{3}$ larger co-volume explored accounting for the magnification in the range of redshift $2.9 <z < 6.7$. On the contrary, the MUSE Wide and other deep blank fields surveys cover well the bright luminosity part \citep[see][]{Drake_2017, Herenz_2019}. These studies are efficient to probe the LF around the $\text{log }L*$ regime, while the sample in this work is optimized to probe the faint-end part. 

For a first attempt, we try to fit the LF with a linear fit to find the slope at the faint end. We use only our LF points, computed from the 575 LAEs. We consider four different redshift bins, 2.9 $< z <$ 4.0 ($z_{35}$), 4.0 $< z <$ 5.0 ($z_{45}$), 5.0 $< z <$ 6.7 ($z_{60}$), 2.9 $< z <$ 6.7 ($z_{all}$) to make these fits. The results are shown in Fig. \ref{fig: LF in 68 percent}. The respective slopes are $-1.76^{+0.26}_{-0.27}$, $-1.99^{+0.12}_{-0.17}$, $-1.96^{+0.5}_{-0.54}$ and $-2.12^{+0.15}_{-0.16}$. In spite of our bigger sample, four times more sources than in \citetalias{Vieuville_2019}, spanning over four orders of magnitudes in galaxy luminosity, we essentially obtain the same results: no evolution of the faint end slope as a function of redshift is observed. The current sample suggests that, at face values, the faint-end slope in each redshift range is steeper than for \citetalias{Vieuville_2019}, for the same redshift interval. However, as seen in Fig. \ref{fig: LF in 68 percent}, a turnover seems to appear at luminosities fainter than $~$10$^{41}$ erg s$^{-1}$, for the two highest redshift bins. This trend is discussed below. 

\begin{table}
\centering
     \caption[]{Best fit parameter values for the Schechter function}
        \label{tab: Best fit parameter values}
        \begin{tabular}{lcccc}
            \hline
            \hline
             Redshift    &$\Phi*$ [10 $^{-4}$ Mpc$^{-3}$]    &log $L*$ [erg s$^{-1}$]  &$\alpha $\\ 
            \hline
            2.9<z<6.7 &7.41$^{+2.70}_{-2.20}$  &42.85$^{+0.10}_{-0.10}$    &$-2.06^{+0.07}_{-0.05}$\\
            2.9<z<4.0   & 6.56$^{+3.20}_{-2.40}$  &42.87$^{+0.11}_{-0.1}$    &$-$2.00$^{+0.07}_{-0.07}$\\
            4.0<z<5.0   &4.06$^{+2.70}_{-1.70}$   &42.97$^{+0.13}_{-0.11}$    &$-$1.97$^{+0.09}_{-0.08}$\\
            5.0<z<6.7   &3.49$^{+2.11}_{-1.50}$ &43.09$^{+0.10}_{-0.08}$    &$-2.28^{+0.12}_{-0.12}$\\
            \hline
        \end{tabular}
  \end{table}
  
\begin{table*}
\centering
     \caption[]{Luminosity bins and LF points used for Fig. \ref{fig: LF in different redshift bin} }
        \label{tab: Luminosity bins and LF points}
        \begin{tabular}{lcccc}
            \hline
            \hline
            \noalign{\smallskip}
       $\text{log }(L)$ [erg s$^{-1}]$& $\text{log }(\phi)(\Delta(\text{log }(L))=1)^{-1} [\text{Mpc}^{-3}]$& $<N>$ & $<N_{corr}>$ &$<V_\text{max}> [\text{Mpc}^{3}]$\\
        \noalign{\smallskip}
        \hline
        2.9<z<6.7\\
        \hline
        39.00<39.63<40.25   &$-0.68^{+0.04}_{-0.35}$& 22.4& 706.0 & 11827\\
        
        40.25<40.38<40.50   &$-0.40^{+0.04}_{-0.39}$& 17.6& 645.4 &15074\\
        
        40.50<40.63<40.75   &$-0.17^{+0.16}_{-0.95}$& 22.8&429.3&28457\\
        
        40.75<40.88<41.00   &$-0.71^{+0.06}_{-0.68}$&45.2 &301.5&31613\\
        
        41.00<41.13<41.25   &$-0.83^{+0.19}_{-0.45}$& 68.9&415.7&37344\\
        
        41.25<41.38<41.50   &$-0.96^{+0.03}_{-0.33}$& 105.0&547.9&41321\\
        
        41.50<41.63<41.75   &$-1.17^{+0.07}_{-0.19}$& 96.4&305.5&42227\\
        
        41.75<41.88<42.00   &$-1.88^{+0.09}_{-0.11}$& 76.4&105.1&46139\\
        
        42.00<42.13<42.25   &$-1.51^{+0.07}_{-0.09}$& 70.4&202.5&45795\\
        
        42.25<42.38<42.50   &$-2.43^{+0.24}_{-0.29}$& 27.5&33.5&47554\\
        
        42.50<42.63<42.75   &$-2.98^{+0.13}_{-0.19}$& 12.9&13.0&49295\\
        
        42.75<42.88<43.00   &$-3.20^{+0.15}_{-0.25}$& 7.7&7.8&49258\\
        
        \hline
        2.9<z<4.0\\
        \hline
        39.00<39.63<40.00 & $-0.15^{+0.07}_{-0.52}$& 6.64&415.33&1712\\
        40.00<40.25<40.50   &$-0.10^{+0.13}_{-0.62}$& 14.19&920.22&6114\\
        
        40.50<40.75<41.00   &$-0.85^{+0.09}_{-0.22}$& 34.0&396.17&11397\\
        
        41.00<41.25<41.50   &$-1.08^{+0.07}_{-0.10}$& 83.7&473.6&14529\\
        
        41.50<41.75<42.00   &$-1.01^{+0.08}_{-0.14}$& 69.5&148.0&15914\\
        
        42.00<42.25<42.50   &$-1.53^{+0.06}_{-0.11}$& 35.6&101.55&16327\\
        
        42.50<42.75<43.00   &$-2.93^{+0.15}_{-0.23}$& 10.0&10.0&17320\\
        \hline
        4.0<z<5.0\\
        \hline
        39.00<39.25<39.50   &$-0.93^{+0.30}_{-1.91}$& 1.0&38.0&730\\
        
        39.50<40.00<40.0&$-1.16^{+0.11}_{-0.22}$& 2.4&48.3&4904\\
        40.0<40.25<40.5 &$-0.38^{+0.09}_{-0.50}$&7.4&311.4&3159\\
        40.5<40.75<41.00   &$-0.38^{+0.11}_{-0.88}$& 19.6&205.1&7662\\
        
        41.00<41.25<41.50   &$-1.43^{+0.10}_{-0.14}$& 51.4&161.2&11044\\
        
        41.50<41.75<42.00   &$-1.48^{+0.1}_{-0.15}$& 55.0&148.5&12164\\
        
        42.00<42.25<42.50   &$-2.30^{+0.13}_{-0.18}$& 30.0&32.2&13182\\
        
        42.50<42.75<43.00   &$-3.15^{+0.20}_{-0.42}$& 4.7&4.8&13433\\
        \hline
        5.0<z<6.7\\
        \hline
        40.00<40.25<40.50   &$-1.55^{+0.20}_{-0.45}$& 6.0&23.8&4725\\
        
        40.50<40.75<41.00   &$-0.89^{+0.24}_{-1.08}$& 14.3&116.5&11105\\
        
        41.00<41.25<41.50   &$-0.66^{+0.15}_{-0.43}$& 38.9&705.5&13545\\
        
        41.50<41.75<42.00   &$-1.52^{+0.19}_{-0.39}$& 48.2&122.9&16190\\
        
        42.00<42.25<42.50   &$-1.63^{+0.09}_{-0.11}$& 32.3&105.2&16705\\
        
        42.50<42.75<43.00   &$-3.19^{+0.2}_{-0.37}$& 5.9&5.9&18542\\

           \noalign{\smallskip}
            \hline
        \end{tabular}
  \end{table*}

 \subsection{Computing LF parameters}\label{sec: LF param}
 
This Section presents the fit of our LF points with the Schechter function. As the luminosity range of our sample reaches its maximum $~$10$^{43}$ erg s$^{-1}$, which is close to values of $\text{log }L*$ (\citeauthor{Herenz_2019} \citeyear{Herenz_2019}, \citetalias{Vieuville_2019}), to completely describe the Schechter function, we need to include other data covering the bright end of the LF. The data included are taken from the works of: \cite{Dawson_2007},
\cite{Blanc_2011},
\cite{Cassata_2011},
\cite{Zheng_2013},
\cite{Sobral_2018},
\cite{Drake_2017}  and
\cite{Herenz_2019}, which have been selected to properly cover the redshift and bright part of luminosity ranges. These data from the literature are averaged with the same luminosity bin size of 0.25 (in log$_{10}(L)$ [erg s$^{-1}$]), except for the last faintest bin which has a width of 1.25 for redshift bin $z_{all}$, while in the other bins the width is 0.5. This is to avoid an increased weight of this bright sample from the literature on the global fit. 

The fitting process is performed as described in Sect. \ref{sec: Schechter fit}. The results are shown in Fig. \ref{fig: LF in different redshift bin} with the best fit curves shown as solid lines together with the 68\% and 95\% confidence area based on the data as indicated in the respective labels. We also check that the shape of the LF for each redshift interval is essentially the same when changing the number of luminosity bins, similarly to \citetalias{Vieuville_2019} (Fig. \ref{fig: LF in different luminosity bin}).

The best fit parameters are listed in Table \ref{tab: Best fit parameter values}. The best fit value of $\text{log } L*$ is in good agreement with \citetalias{Vieuville_2019}, and a few percents higher than the value obtained from \cite{Herenz_2019}, 42.20$^{+0.22}_{-0.16}$. Strong degeneracy between these parameters ($\Phi*$, $L*$, $\alpha$) is observed, as shown in Fig. \ref{fig: 3 parameters correlation}, and already well documented in \cite{Herenz_2019}. $\text{log }L*$  seems to be well measured from the current work for different redshift intervals. There is a tendency of $L_*$ to increase with redshift but this is well within the uncertainties. As both $\text{log }L*$ and the steep faint-end slope increase with redshift, it is likely related to the degeneracy mentioned earlier. 

$\Phi*$ is just a normalization factor giving the number density of objects per given volume. Our best fit result gives \mbox{$\Phi*$ [10$^{-4}$\text{Mpc$^{-3}$}] = $7.41^{+2.70}_{-2.20}$} which is consistent with \citetalias{Vieuville_2019} and \cite{Sobral_2018}, but smaller than that of \cite{Herenz_2019}. The $\Phi*$ value strongly depends on the literature data points used for the fitting procedure.

As our sample probes the faint luminosity regime, the slope at the faint-end of LF, $\alpha$, in principle, is well constrained. We measure steep slopes of $\alpha$ varying from $-$2.0$\pm$0.07 for the redshift interval $2.9<z<4.0$ to $-$2.28$\pm$0.12 for redshift interval  $5.0<z<6.7$. These results are consistent with the slope measured by \cite{Herenz_2019} for the global redshift bin, and \cite{Drake_2017} in same redshift bins. The faintest luminosity points in all redshift intervals, having log($L$) [erg s$^{-1}$] < 40, are not included in the Schechter's fitting. As sources in this faintest bin are often highly magnified by lensing effects, i.e. sources close to the lens caustic lines, this seems to suggest either that lensing systematic uncertainties or that completeness corrections for those sources must be treated with great care. Another possibility is that the faint end of the LF may depart from the traditional Schechter function. More data in this luminosity regime is required to verify this. The enhancement at log(L) around 42 shown in $z_{all}$ is because the number of sources suddenly increases in that luminosity bin. This is also shown in Fig. \ref{Fig: Source_distribution}, where a spike appears at the luminosity bin log($L$) [erg s$^{-1}$] $\sim$ 42 after correcting for completeness. This may relate to the over-density of background sources at $z\sim4$ as mention in \citepalias{Vieuville_2019}. This may also suggest the uncertainty from the cosmic variance is probably larger than expectations. The co-volume probed by our survey is $\sim$ 50,000 \text{Mpc}$^3$. It seems that the data points from \cite{Drake_2017} and \cite{Cassata_2011} within the log(L) range of 41.5-43.0 might influence the fitting results of the faint end slope $z_{35}$ and $z_{60}$. However, we have checked and found that including or excluding these data points does not substantially alter our results. The faint end slopes at these specific redshift ranges show only minor variations less than a few percent.

\subsection{Error budget}

The uncertainties on LF points are well documented in \citetalias{Vieuville_2019}. There are three kinds of uncertainties attached to the LF points which are addressed. The statistical uncertainty is derived from the MCMC process. A set of flux, magnification and completeness associated with each LAE is randomly drawn, 20,000 times. The results are then used to estimate the mean and statistical uncertainties attached to each LF point. The second source of uncertainty relates to field $-$ to $-$ field variance for different lensing Fields of View observed by MUSE. We use the cosmic variance calculator proposed by \cite{Trenti_2008} to estimate this uncertainty, which is typically about 20\% to 30\% at most. The third one, the Poissonian uncertainty, relates to the number of sources for a given luminosity bin, which is relatively easy to handle. For the bright-end luminosity, log($L$) [erg s$^{-1}$] > 42, the Poissonian uncertainty dominates. Its contribution decreases and becomes equivalent to that of cosmic variance in the luminosity range 41 < log($L$) [erg s$^{-1}$] < 42. The statistical uncertainty is dominant in the faint-end regime.

One would expect systematic uncertainty coming from lensing models to be another important contribution to the total error budget. Different lensing models may give magnification factors that differ by a significant factor for a given source, making larger the uncertainty of the luminosity obtained.  This systematic uncertainty is well documented in \cite{Bouwens_2017}, \cite{Atek_2018}, \cite{Priewe_2017}, \cite{Menegetti_2017}. It might play important role in particular to the faint end luminosity regime. As discussed in \citetalias{Vieuville_2019} its contribution is $\sim$15\% at log(L) of 40.5 erg s$^{-1}$ for the case of the lensing field A2744.

 \section{Discussion}\label{sec: 6}

 \subsection{Uncertainties associated to the computation of the LF in strong lensing fields} 

Thanks to the magnification by lensing clusters, we can reach a fainter galaxy population compared to that of \cite{Drake_2017} and \cite{Herenz_2019} by one order of magnitude, allowing us to constrain the faint-end slope of the LF. Although the number of sources in the previous sample by \citetalias{Vieuville_2019} is four times smaller, it still covered the same luminosity range as in the present project. Three lensing clusters, which were previously studied in \citetalias{Vieuville_2019}, 
have been re-reduced to achieve better quality in terms of S/N and source positions. 

In this work, when sources are both detected by HST and MUSE, we use the Muselet positions (M pos.) instead of prior ones from HST observations (P pos.). Choosing these positions has some impact on the LF shape at the faint end. An average spatial offset of $\sim$0.2 arcsec between the M and P positions has been found in \cite{Claeyssens_2022}. MUSE observations are sensitive to detect line emissions such as Lyman-alpha while HST ones are sensitive to continuum emission. It may happen that the LAEs are extended and associated with diffuse gas, hence the offset between the two observations. 

Moreover, the faint luminosity regime often contains  highly magnified sources. These sources are close to the lens caustic lines in the source plane, hence their positions have to be assessed with great care. The weighted magnifications for these extended sources strongly depend on their positions with respect to the caustic lines. Not to mention that the magnification uncertainties attached to these sources are also large, and this may cause the source contribution to spread to several luminosity bins during the MCMC process to estimate LF point uncertainties. On the contrary, sources in the bright luminosity bins usually have small magnifications and do not suffer from this effect. In principle, the procedure adopted here to compute the LF points captures all these effects. 

An important point to note: \citetalias{Vieuville_2019} used a threshold of 10$\%$ completeness to reject sources having completeness values below this cut. A small completeness value  implies a large correction on the number of detected sources (see Fig. \ref{Fig: Source_distribution}). Varying the completeness threshold cut significantly changes the shape of the LF at the faint end as one may loose some fraction of these sources. In the present work, we try to include as many sources as possible from the sample, only rejecting obvious cases which have a poor extraction quality, i.e. flag type 3 as identified from SExtractor. In practice, we use a cut in completeness of 1\%, meaning that a final sample of 575 LAEs is used for the LF computation. In the global redshift bin, $z_{all}$, and $z_{35}$ (see Fig. \ref{fig: LF in different redshift bin}) the number density of sources suddenly increases at log ($L$) [erg s$^{-1}$] $\sim$ 42, this is also caused by the low completeness values for some sources in that luminosity bin. It is worth noting that usual computations of the LF in lensing fields do not reject any source based on its completeness value \citep[see  e.g. ][]{Atek_2018}. By implementing a 10\% completeness, we further remove 62 additional sources from our sample, with 51 of them belonging to the six faintest luminosity bins. As a result, the density of source significantly reduces by a factor of 5 on average. The faint end slope decreases from its original value of $-$2.06 to a flatter value of $-$1.46 for the global redshift range Table \ref{tab: different slope value}. This reveals insights into the uncertainties associated with the faint end slope, which we will discuss in detail in the subsequent subsection.
 
\subsection{Comparison with previous results}

In the bright luminosity regime (\text{log }($L$) [erg s$^{-1}$] > 40.5), the data from the literature mainly come from blank field observations. They are numerous and helpful to constrain the LF at the bright end. Indeed, we include some of them from blank fields for the fitting procedure to constraint the bright-end part of the LF. \cite{Blanc_2011} used 89 LAEs with a redshift range 1.9 $<z<$ 3.8 obtained from Hobby Eberly Telescope Dark Energy Experiment Pilot Survey (HETDEX) to study the LF using the same $V_\text{max}$ method. \cite{Drake_2017} used 604 LAEs in the redshift range 2.91 $<z<$ 6.64 obtained from VLT/MUSE. \cite{Cassata_2011}  used 217 LAEs in the redshift range 2 $<z<$ 6.62 obtained from Vimos-VLT Deep Survey. \cite{Sobral_2018} studied $\sim$4000 LAEs from $z\sim$ 2 to 6 covering a luminosity range of 42.4 < \text{log }($L$) [erg s$^{-1}$] < 43.0 obtained from the Subaru and the Isaac Newton Telescope in the $\sim 2 \text{deg}^2$ COSMOS field. Our LF points in the bright part (log($L$) [erg s$^{-1}$] > 42) are consistent with these results from the literature (Fig. \ref{fig: LF in different redshift bin}). 

It is worth mentioning the work of \cite{Herenz_2019}. They used data including 237 LAEs from the MUSE Wide survey to construct the LF in the same redshift ranges as this work. They found the faint-end slope value at redshift 2.9 < $z$ < 6.7 to be $\alpha=-1.84^{+0.42}_{-0.41}$, log $\Phi*$ [{Mpc}$^{-3}]=-$2.71 and \mbox{log $L*$[erg\ s$^{-1}$]=42.20$^{+0.22}_{+0.16}$}. Comparing to our results, the two slopes at the faint end are consistent while their best fit of $\text{log} L*$ is a bit smaller. The explanation may relate to the degeneracy between the three best fit parameters. We note that \cite{Herenz_2019} constructed LAEs LF in the luminosity range 42.2 < \text{log }($L$) [erg\ s$^{-1}$] < 43.5.

It is necessary to compare our results with those obtained by \citetalias{Vieuville_2019},
which also probe the faint luminosity regime. The characteristic $\text{log } L*$ is well measured both in different redshift intervals and in different cluster samples. The best-fit values agree well with each other, within their 1$-\sigma$ uncertainties. However, the present slope values are steeper by 20\% than those from the previous work for the same redshift intervals. The difference between the two may be due to various factors. Firstly, the number of sources in the two samples differ: we have four times more sources in the present sample, giving us better statistics. Our sample has a significant number of sources in the faint luminosity region and more than 10 sources in the faintest bin. Secondly, the threshold cuts in  completeness are different. \citetalias{Vieuville_2019} rejected faint sources with completeness values below 10\%, while we have included as many sources as possible, for consistency with other LF determinations in lensing fields, rejecting only those with completeness values below 1\%. Thirdly, as explained above, our MCMC procedure to compute the LF points better captures the relationship between magnification and $V_\text{max}$ with respect to \citetalias{Vieuville_2019}. 
Finally, the number of LF literature data points helping to constraint the bright-end part is different between the two works, ie. The results from \cite{Herenz_2019} using MUSE-Wide survey to investigate Lyman $\alpha$ LF in the same redshift bin are included and combined with others as a constraint association; the results from \cite{Drake_2017} are used for fitting at redshift bin $2.9 <z< 4.0$ and for displaying only at others bins. This affects the normalization parameter as well as the faint-end slope, as the two are correlated. We have also performed the full analysis with the new improved procedure  on the sub-sample of four clusters in \citetalias{Vieuville_2019}, with the same choices regarding the completeness, and found fully consistent results with \citetalias{Vieuville_2019}.
  
One of the important points derived from Fig. \ref{fig: LF in different redshift bin} is that faint-end slope values of $\alpha$ become steeper at higher redshifts. This seems to suggest the evolution of the faint-end slope as a function of redshift. There is a very good agreement between the slopes obtained from Schechter function fitting and those obtained from line-fitting in Section \ref{sec: 5.1}. However, we do not see the same trend of the slope evolution from line-fitting, which only used our LF points, due to the larger uncertainties associated with them. Moreover, we observe a turnover at luminosities fainter than $~$10$^{41}$ erg s$^{-1}$, for the two highest redshift bins. At these luminosities, the LF points are not well described by the Schechter function. A flattening/turnover is observed towards the faint end for the highest redshift bins that still needs further investigation. This turnover is similar to the one observed at ${M_{UV} = -15}$ for the UV LF at z $\ge$ 6 in lensing clusters, with the same conclusions regarding the reliability of the current results when we compare, for instance, the work of  \cite{Atek_2018} to \cite{Bouwens_2022} (see discussion below). 

To account for a possible turnover in the two highest redshift ranges 4.0 < $z$ < 5.0 and 5.0 < $z$ < 6.7, two more parameters have been introduced to modify the Schechter function, which now reads:
\begin{equation}
\Phi(L)\text{ exp}{\{-(L_T/L)^m\}}=\frac{\Phi*}{L*}\left(\frac{L}{L*}\right)^\alpha \text{ exp}{(-L/L*)} \text{exp}{\{-(L_T/L)^m\}}
\end{equation}
where: $\Phi(L)$ is the traditional Schechter function together with its three ($\Phi*, L*$, $\alpha$) parameters, being fixed at their best-fit values, as shown in Table \ref{tab: Best fit parameter values}. Two new parameters are introduced: $L_T$ is the luminosity at the turnover point and $m$ is a power index, which is about unity.
Our data suggest that log($L_T$) are $\sim$40 and $\sim$40.7 erg s$^{-1}$ for redshift bins 4.0 < $z$ < 5 and 5.0 < $z$ < 6.7, respectively. The values of $m$ are essentially equal to 1 for these two redshift ranges, as expected. Results are shown in Fig. \ref{fig: LF in different redshift bin} (lower panels).

Regarding the prevalence of a turnover in the LF towards the faint end, it is worth mentioning that \cite{Bouwens_2022} have recently ruled out this trend in the UV LF down to $M_{UV} = -14$ mag at $z\sim$6. In addition, \cite{Dawoodbhoy_2023}, using the CODA simulation, report no sign of the turnover down to ${M_{UV} = -12}$ mag at the same redshifts. The large uncertainty in the faint regime of LAE luminosity prevents us to make much sense of this result. Additional data in this faint region are necessary to improve the statistical significance of the present sample in order to confirm/reject this turnover trend.

\subsection{Comparison with theoretical predictions}

\begin{figure}[!h]
   \centering
   \includegraphics[width=0.5\textwidth]{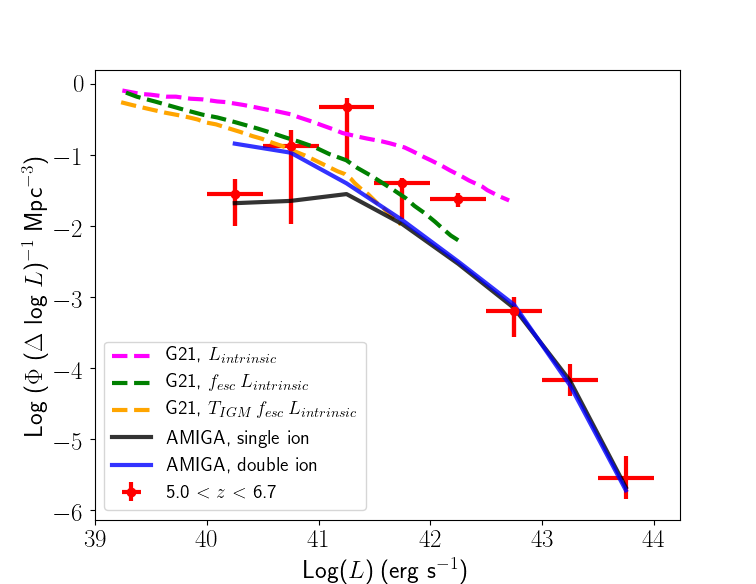}
   \caption{Comparison with model predictions. Red crosses show our reconstructed LF points in the redshift range of 5.0 < $z$ < 6.7. Two additional crosses in the bright end, having log($L$)[erg s$^{-1}$]$>$43, are taken from the literature, obtained the same way as described in Sec. \ref{sec: LF param}. Black and blue solid lines show predictions from the AMIGA models with single- and double-ionization scenarios, respectively. Predicted LFs from the SPHINX simulation \citeauthor{Garel_2021} \citeyear{Garel_2021}  are shown as dash lines: intrinsic LF (magenta), attenuation by dust (green) and by IGM (orange).}
   \label{fig: compamiga}
    \end{figure}

We have compared our results on the Lyman-alpha LF with two theoretical models whose predictions at $z\sim 6$ can be compared directly with our findings in the highest redshift range of 5.0 $ < z < $6.7, namely the predictions by \cite{Garel_2021} and by \cite{Salvador_2022}.
The first one \citep{Garel_2021} predicts the Lyman-alpha LF at the epoch of reionization by computing the radiative transfer of Lyman-alpha from ISM to IGM scales, using the SPHINX radiation-hydrodynamics cosmological simulation. Their results, obtained by computing the intrinsic LF function, and the attenuation by dust and then by IGM, are shown in Fig. \ref{fig: compamiga}. 
The second one \citep{Salvador_2022} uses the Analytic Model of Igm and GAlaxy evolution (AMIGA), a model of galaxy formation and evolution including their feedback on the IGM, to constrain the reionization history of the universe. It provides predictions for two possible scenarios: single and double reionization episodes. The former has hydrogen ionization episode at $z\sim$6, and the latter has two reionization episodes at $z\sim$6 and $\sim$10, separated by a short recombination period \citep[see][for additional details]{Salvador_2022}. Their results for the two scenarios are also shown in Fig. \ref{fig: compamiga}. In the range of 40<log($L$)[erg s$^{-1}$]<42, the prediction of the AMIGA double ionization scenario is in good agreement with that of \cite{Garel_2021} after IGM correction. Our LF points towards the faint end correctly span the region covered by these models. In general, our LF points are in good agreement with the predictions by both models without any renormalisation, as shown in Fig. \ref{fig: compamiga}. At the faintest luminosity regime, log($L$)<41, our LF point starts to depart from the theoretical increasing trend by \cite{Garel_2021}, and is somewhat closer  to the single ionization scenario predicted by \cite{Salvador_2022}.
However, the uncertainties in this regime are very large, preventing us to distinguish between the different theoretical predictions. More data covering this faint regime are badly needed. 

\subsection{Effect of source selection}
There are several factors that play important roles in determining the faint end slope. One of them is the source selection process. In the 17 lensing clusters, 190 LAEs (unique system) have been classified as $zconf=1$, and as a consequence, they are not retained for the LF computation process. However, as they are often faint, they may contribute significantly to our faint LF points. To evaluate the impact of $zconf=1$ sources on the final LF points, we have incorporated them into our LAE sample. We assume that $zconf=1$ sources have the same quality of completeness and $V_{\text{max}}$ as the $zconf=2,3$ sources. Namely, for a given luminosity bin, $zconf=1$ completeness and $V_{\text{max}}$ values are assigned as the mean values for the $zconf=2,3$. Similarly, the uncertainty values for completeness and $V_{\text{max}}$ of the $zconf=1$ sources are set equal to the corresponding uncertainty values of the $zconf=2,3$ sources in that bin. Including half and all of the $zconf=1$ sources results in a 5\% and 10\% steeper faint end slope, respectively.

\begin{figure}[!h]
   \centering
   \centering
   \includegraphics[width=0.5\textwidth]{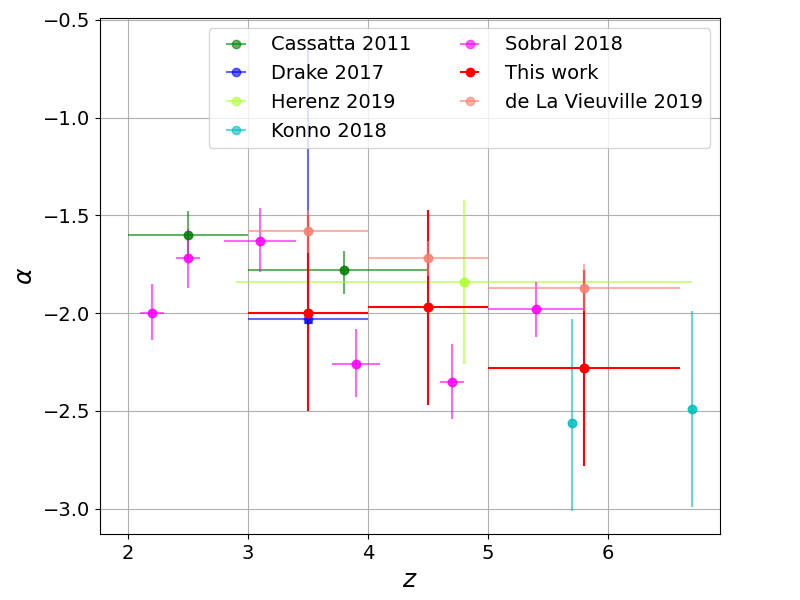}
   \caption{Comparison of slope evolution with redshift from this work (red crosses) to the literature which appears in the Fig \ref{fig: LF in different redshift bin}. The horizontal error bars are the redshift range of the surveys. }
   \label{fig: alpha_comparison}
    \end{figure}

To estimate the systematic uncertainties attached to the faint end slope, we performed various tests to calculate the LF points and measured the slope. We employed different completeness threshold cuts (1\% and 10\%), ultilized different fitting function forms (Schechter and linear functions), and considered two senariors: including half or all of the $zconf=1$ sources. We also used source fluxes obtained from the pipeline only (method 2). Our findings are summarized in Table \ref{tab: different slope value}. We retain faint end slope of the 1\% completeness cut results and enlarge the uncertainties. The final slope are $-2.00\pm0.50, -1.97\pm0.50, -2.28\pm0.50, -2.06\pm0.60$ for $z_{35}, z_{45}, z_{60}, z_{all}$, respectively. Fig. \ref{fig: alpha_comparison} illustrates the faint end slope at different redshift ranges. The slope shows a slight increase with redshift, although the uncertainties remain large. Our slopes at various redshift ranges are in good agreement with the findings of other studies, typically within a 1$-\sigma$ deviation.
    
\begin{table*}
\centering
     \caption[]{Results of the faint end slope $\alpha$ from different tests}
        \label{tab: different slope value}
        \begin{tabular}{lcccc}
            \hline
            \hline
          & $z_{35}$& $z_{45}$& $z_{60}$ & $z_{all}$    \\ 
          \hline
          Schechter fitting&  & & &\\
          \hline
          1\% completeness cut & $-2.00\pm 0.07$ & $-1.97\pm0.09$ & $-2.28\pm0.12$ & $-2.06\pm0.07$\\
          10\% completeness cut & $-1.78\pm0.06$ & $-1.83\pm0.12$ & $-1.75\pm0.10$ & $-1.46\pm0.05$\\
          1\% completeness cut, fluxes obtained from the second method & $-2.10\pm0.06$ & $-1.97\pm0.08$ &
          $-2.24\pm0.11$ &
          $-1.82\pm0.03$\\
          1\% completeness cut zconf1 included & $-1.83\pm0.17$ & $-1.92\pm0.09$ &
          $-1.94\pm0.10$ &
          $-2.29\pm0.09$\\
          \hline
          Linear fit\\
            \hline
            1\% completeness cut & $-1.76\pm0.27$ &
            $-1.99\pm0.17$ &
            $-1.96\pm0.54$ &
            $-2.12\pm0.16$\\
            10\% completeness cut & $-1.55\pm0.17$ &
            $-1.63\pm0.21$ &
            $-1.66\pm0.24$ &
            $-1.64\pm0.12$\\
            Faint end slope retain &
            $-2.00\pm0.50$ &
            $-1.97\pm0.50$ &
            $-2.28\pm 0.50$ &
            $-2.06\pm0.60$\\
            \hline
        \end{tabular}
     
  \end{table*}

\subsection{Implications for the reionization}

 One of the most important tools to understand the formation and evolution of galaxies is the cosmic star formation rate density (SFRD), providing information on the onset of star formation in the early Universe and its evolution over cosmic time \citep{Madau_1996, Bouwens_2007, Bouwens_2008}. In this section, we present the contribution of LAE population to the cosmic re-ionization by computing the SFRD by integrating the best fit parameters obtained from the previous section. In addition to the luminosity range of the integral (Eq. \ref{func: SFRD}), log($L$)=(39.5, 44.0) as probed by the present sample, we also choose another lower limit luminosity of 0.03$L*$ $\sim$10$^{41}$ erg s$^{-1}$ to facilitate the comparison with previous works.  The integrated SFRD is proportional to the luminosity density and can be estimated following the calibration of \cite{Kennicut_1998}, assuming an intrinsic factor of 8.7 between the intrinsic $Ly\alpha$ and $H\alpha$ fluxes, and case B for the recombination \citep{Osterbrock_1989}. In that case, all newly formed $Ly\alpha$ photons would be re-absorbed by the neutral hydrogen atoms of the HII region. In the optically thick case, the SFRD is written as:
  \begin{equation}
      SFRD_{Lya}[\text{M}_{\odot} \text{yr}^{-1}\text{Mpc}^{-3}] =\rho_{Lya}/1.05 \times 10^{42}
      \label{func: SFRD}
  \end{equation}
  where $\rho_{Lya}$ is Lyman alpha luminosity density in units of \mbox{erg s$^{-1}$ Mpc$^{-3}$}.

\begin{figure}[!h]
   \centering
   \centering
   \hspace{-0.6 cm}
   \includegraphics[width=0.5\textwidth]{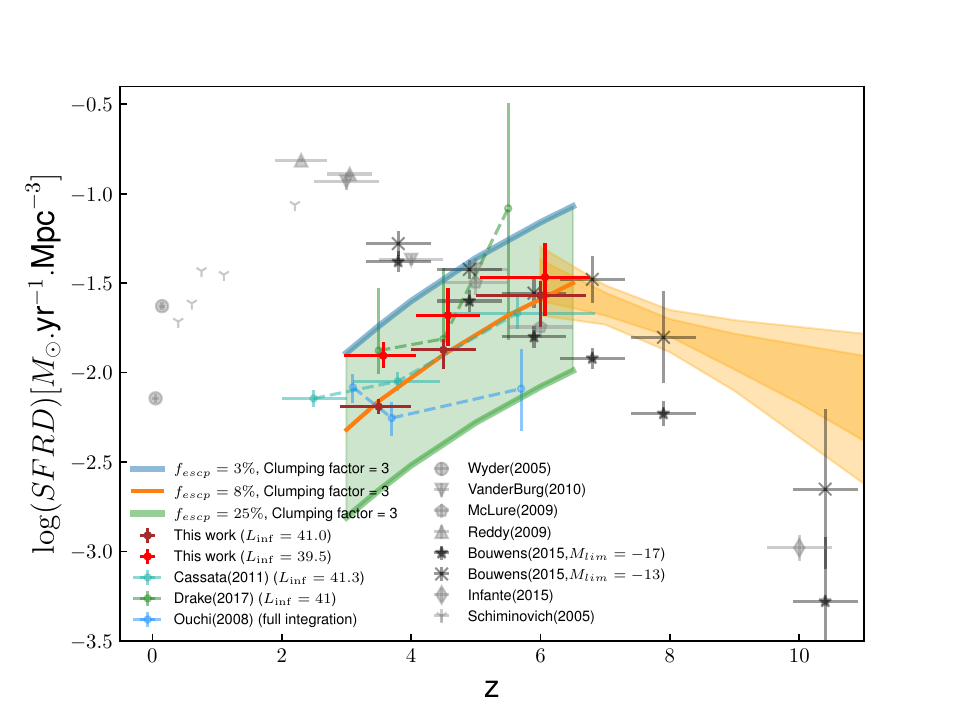}
   \caption{Evolution of the star formation rate density with redshift using different integration limits for the LF. Brown (red) crosses show results of integrating the best-fit Schechter (modified Schechter) functions log($L$) from 41 (39.5) to 44 erg s$^{-1}$. Solid  blue, orange and green curves indicate the critical SFRDs obtained by assuming a clumping factor of 3 and $f_{escp}$= 3$\%$, 8$\%$ and 25$\%$, respectively. The solid green curve, having $f_{escp}$= 8$\%$, indicates the best likely match with our SFRD points in the luminosity range of 41<log($L$)<44 (see text).
   }
   \label{fig: sfrd evolution}
    \end{figure}

Results are shown in Fig. \ref{fig: sfrd evolution} together with the other literature data from different sources/surveys. The yellow regions show the SFRD needed to fully ionize the entire Universe, taken from the work of \cite{Bouwens_2015b} at the level of 1$-\sigma$ and 2$-\sigma$. A clumping factor value of 3 is applied to calculate cosmic emissivity,  log$(\xi_{ion}f_{escp})$=24.50, where $f_{escp}$ is escape fraction of ionizing UV photons, $\xi_{ion}$ is the production efficiency of Lyman-continuum photons per unit UV luminosity. The conversion to SFRD is then calculated as, $SFRD_{Lya}[M_{\odot} yr^{-1} Mpc^{-3}] =\rho_{UV}/(8.0 \times 10^{27})$.

Fig.\ref{fig: sfrd evolution} shows the  evolution of the SFRD with  redshift. As the LF decreases steeply toward the bright-end, the upper limit of the integration (Eq. \ref{func: SFRD}) does not affect to the final result. However, the steep slope at the faint-end plays an important role. If one takes the integral over the full range probed by the present sample, the contribution of LAEs to the cosmic re-ionization at redshift $z\sim 3.5$ and $z\sim 4.5$ would be $\sim 80\%$ compared to that of \cite{Bouwens_2015b} using $M_{UV}$=$-$17 mag as observational limit. At higher redshifts $z\sim 6$ the contribution is much higher, 10\% larger than \cite{Drake_2017}, 4 times larger than \cite{Cassata_2011}, 12 times larger than \cite{Ouchi_2008}, and much higher than \cite{Bouwens_2015a}. If one takes the lower limit of the integral (Eq. \ref{func: SFRD}) at 0.03 $L*$, results are shown as reddish-brown crosses, which are more consistent with others mentioned above. Then, the contribution of LAEs population at redshift $z\sim 3.5$ is $10\%$, $30\%$ at redshift $z\sim 4.5$, and reaches up $100\%$ at higher redshifts. Fig. \ref{fig: sfrd evolution} also shows SFRD reported by \cite{Sobral_2018}. The difference between our results on the evolution of the SFRD and those obtained by \cite{Sobral_2018} can be explained by comparing the detailed shape of the LF and the luminosity range covered by the two studies. Indeed, the shape of the bright end does not affect the SFRD, and our sample better captures the steepening of the LF towards the faint end, which is responsible for the increase of the SFRD with redshift in our case.
When using the modified Schechter functions instead of the traditional one to compute the SFRD over the full probed luminosity range, 39.5 < log($L$) < 44, in order to better describe the flattening/turnover observed towards the faint end, the contribution is 25$\%$, 50$\%$ and 100$\%$, respectively. 

Assessing the ability of a population of sources to 
reionize the Universe is usually done by comparing its ionizing power to the critical value needed to maintain reionization at a given redshift. This value for the critical photon emission rate per unit cosmological comoving volume was introduced by \cite{Madau_1999} as follows:
\begin{equation}
   \Dot{N}=(10^{51.2}\text{s}^{-1} \text{Mpc}^{-3}) {C_{30}}\left(\frac{1+z}{6}\right)^{3} \left(\frac{\Omega_{b}h^2}{0.08}\right)^{2}
    \end{equation}
where $C_{30}$ is the clumping factor $C_{HII}$=<$n^2_{HII}$>/<$n_{HII}$>$^2$, normalized to $C_{HII}$=30, and $n_{HII}$ is the mean comoving hydrogen density in the Universe. Assuming $C_{30}=1$ and $\Omega_{b} h^2_{50}=0.08$, the critical  SFRD will be written as: 
\begin{equation}
    SFRD_{crit}=\Dot{N}\times 10^{-53.1} f^{-1}_{escp}\sim 0.013 f^{-1}_{escp} \times \left(\frac{1+z}{6}\right)^{3} 
    \end{equation}
The clumping factor, $C_{30}$ can be considered as a correction accounting for  inhomogeneities in the IGM, and it is supposed to vary  as a function of redshift. For example, as reported by \cite{Shull_2012}, $C_{HII} = 2.9 \times ((1+z)/6)^{-1.1}$, namely a value $\sim$3 at $z\sim$5 hence the critical SFRD obtained from the equation above needs a correction factor of $\frac{C_{HII}}{30}$. Here we adopt the average value of $C_{HII} = 3$, to facilitate the comparison with previous works  
\citep{Pawlik_2009, Robertson_2013, Robertson_2015, Bouwens_2015b, Gorce_2018}.

The dependence of Ly$\alpha$ escape fraction on redshift is an important quantity as it helps to constrain the reionization history of the Universe. The Ly$\alpha$ escape fraction $f_{escp}^{Ly\alpha}$ and $f_{escp}$ are expected to be correlated \citep[see e.g.][]{2016ApJ...828...71D,2020MNRAS.491..468I}, \cite{Hayes_2011} reported an evolution as a function of redshift as follows $f_{escp}^{Ly\alpha}$ $\propto (1+z)^{\xi}$, with $\xi$ = 2.57$^{+0.19}_{-0.12}$ over the redshift range 0.3 < $z$ <6, reaching a  maximum of unity at $z$=11.1. The increase in $f_{escp}^{Ly\alpha}$ with redshift follows the evolution of the dust content in galaxies up to $z \sim$ 6, and then drops above $z \sim$ 6.5. Using this prescription in the range covered by our study and a reasonable conversion for $f_{escp}$ following \cite{2016ApJ...828...71D}, the order of magnitude expected for 
$f_{escp}$ is $\sim$5$\%$ at $z\sim$ 3 and up to some 25$\%$ at $z\sim$ 6. Fig. \ref{fig: sfrd evolution} displays the critical SFRDs obtained from these two extremes of $f_{escp}$ as a shaded area, using the same clumping factor of 3 as described above. As seen in the figure, the SFRD points obtained from our Ly$\alpha$ LF with the integration limits 41<log ($L$)<44 are fully consistent with the critical ones for the average value of $f_{escp}\sim$8$\%$ and $C_{HII} = 3$, in the redshift interval 3 < $z$ < 6.7. In other words, the contribution of LAEs to the ionizing flux in this redshift interval seems to be sufficient to keep the hydrogen ionized. The contribution of LAEs at $z \sim 6$ is comparable to the one provided by LBGs. 

This result taken at face values suggests that the contribution of the population of LAEs to  cosmic re-ionization could be much higher than previously expected. There are, however, a few caveats to mention.  Firstly, the vast majority of previous surveys in blank fields have been focused on the most luminous galaxies in their restframe UV, that are well suited for a successful spectroscopic follow up. By construction, these samples can not probe to the faint luminosity regime, down to $10^{39}$ erg s$^{-1}$, as in the present work. In addition, IFU observations identify LAEs without any pre-selection. Secondly, the contribution to the SFRD is directly related to the steep slope value being obtained, which is directly proportion to Ly$\alpha$ luminosity density. A 20\% steeper slope dramatically changes the LAEs contribution to the cosmic re-ionization as illustrated here. Thirdly, the lower limit of the integration also has an impact on the final contribution. Moreover, there are still large uncertainties in the Ly$\alpha$ $f_{escp}$ and its evolution with redshift making difficult the comparison between different works using different prescriptions.  

\section{Conclusions}\label{sec: 7}
 
We have presented a study of the galaxy Ly$\alpha$ LF using a large sample of 17 lensing clusters observed by MUSE. We blindly selected 600 lensed LAEs behind these clusters in the redshift range 2.9 < $z$ < 6.7. The sample covers four orders of magnitude in galaxy luminosity ($39.0<\text{log }(L)<44.0$) and probes efficiently to the faint luminosity regime, down to 10$^{40}$\ erg\ s$^{-1}$. This sample sets a strong constraint on the LF at the faint end as well as the evolution of the slope as a function of redshift. 

To deal with the combination of both lensing fields and spectroscopic datacubes obtained from MUSE, we adopt the same V$_\text{max}$ method and the same procedure as described in \citetalias{Vieuville_2019} to compute the LF points. Several improvements have been introduced and applied to the original pipeline, allowing us to better account for lensing  magnification. The main results and conclusions are listed as follows:

- Regarding the method and the pipeline, several effects have been studied in details, with the corresponding improvement on the final results with respect to \citetalias{Vieuville_2019}. The new pipeline takes better into account the magnification errors when computing V$_\text{max}$. A careful analysis has been performed on the effects of source centering and completeness corrections. 

- We have studied the LF in four redshift bins, $2.9<z<4.0$, $4.0<z<5.0$, $5.0<z<6.7$, and $2.9<z<6.7$. The total co-moving volume of our survey is $\sim$50,000 Mpc$^3$. With respect to the previous work by \citetalias{Vieuville_2019}, A2744 is still the  dominant cluster, with a  contribution to the total volume three times above the average. 

- The three best fit parameters of the Schechter function in the redshift range 2.9 < $z$ < 6.7, obtained from our sample, are: $\alpha=-2.06^{+0.07}_{-0.05}$, $\Phi*$[10$^{-4}$ Mpc$^{-3}$] = $7.41^{+2.70}_{-2.20}$, log($L*$)\text{[erg s$^{-1}$]} = $42.85^{+0.10}_{-0.10}$.

- The LF values in the bright-end regime (logL > 42) are consistent with previous works, in particular the MUSE-Wide observations, as well as with other surveys in the literature using different techniques.

- In the faint luminosity regime, the contribution of highly magnified sources to the LF points in the faint luminosity bins is significant, as expected. The density of sources is well described by a steep slope, $\alpha\sim-2$. The two slopes obtained from line-fitting and from the Schechter function are consistent within their uncertainties. The present Schechter slope is consistent with those from \cite{Drake_2017} for each redshift interval, with \cite{Herenz_2019} for the global redshift range, and is $20\%$ steeper than the slope of \citetalias{Vieuville_2019}. When taking into account different flux measurements, different completeness threshold cut (1\% and 10\%), and source selection effects, we obtained faint end slopes of $-2.00\pm0.50, -1.97\pm0.50, -2.28\pm0.50 \text{,and} -2.06\pm0.60$ for $z_{35}, z_{45}, z_{60}, \text{and} z_{all}$, respectively. These results are in line with other studies, typically within a $1-\sigma$ deviation. 

- The steepening of the faint end slope with redshift, suggested by the earlier work of \citetalias{Vieuville_2019} is confirmed, but the uncertainties remain large. A turnover seems to appear at luminosities fainter than $\sim$10$^{41}$ erg s$^{-1}$, for the two highest redshift bins. 

- The SFRD depends strongly on the interval of luminosity over which the density is integrated. The steeper slope at the faint end causes the SFRD to dramatically increase between $z \sim $ 3 and 6, implying that LAEs play a major role in the process of cosmic reionization.

- The contribution of LAEs to the ionizing flux in the redshift interval studied here seems to be sufficient to keep the hydrogen ionized. The contribution of LAEs at $z\sim $ 6 is comparable to the one provided by LBGs. 

\begin{acknowledgements}
We are grateful to the referee for a very careful reading of the manuscript and pertinent comments that helped greatly with improving the substance of the present article. 
     This work is done based on observations made with ESO Telescopes
at the La Silla Paranal Observatory under programme IDs 060.A-9345,
094.A-0115, 095.A-0181, 096.A-0710, 097.A0269, 100.A-0249, and
294.A-5032. Also based on observations obtained with the NASA/ESA Hubble
Space Telescope, retrieved from the Mikulski Archive for Space Telescopes
(MAST) at the Space Telescope Science Institute (STScI). STScI is operated by
the Association of Universities for Research in Astronomy, Inc. under NASA
contract NAS 5-26555. This research made use of Astropy, a community-
developed core Python package for Astronomy (Astropy Collaboration 2013).
All plots in this paper were created using Matplotlib (Hunter 2007). Financial support from the World
Laboratory, the Odon Vallet Foundation and VNSC is
gratefully acknowledged. Tran Thi Thai was funded by Vingroup JSC and supported by the Master, PhD Scholarship Programme of Vingroup Innovation Foundation (VINIF), Institute of Big Data, code VINIF.2022.TS.107. This work received support from the French government under the France 2030 investment plan, as part of the Excellence Initiative of Aix-Marseille University - A*MIDEX (AMX-19-IET-008 - IPhU). 
\end{acknowledgements}
\bibliographystyle{aa}
\bibliography{aa.bib}

\begin{appendix} 
\clearpage
\onecolumn
\section{Lens models of 17 clusters}
    \begin{longtable}{ccccccccl}
    \label{table: lens model}\\
    \caption{Best fit parameters of mass distribution in each cluster}\\
    \hline
    \hline
       Cluster & $\Delta$RA["] & $\Delta$DEC ["] & $\epsilon$ & $\theta$[deg] & 
       $r_{core}[kpc]$& $r_{cut}$[kpc] & $\sigma $[km $s^{-1}$]& Reference \\
        \noalign{\smallskip}
        \hline
        \endfirsthead
        \caption{continued.}\\
        \hline\hline
        Cluster & $\Delta$RA["] & $\Delta$DEC ["] & $\epsilon$ & $\theta$[deg] & 
       $r_{core}[kpc]$& $r_{cut}$[kpc] & $\sigma $[km $s^{-1}$]& Reference \\
       \hline
       \endhead
       \hline
       \endfoot
     A2390\\
      \hline
DM1 &$31.6^{+1.8}_{-1.3}$&$15.4^{+0.4}_{-1.0}$&$0.66^{+0.03}_{-0.02}$& $214.7^{+0.5}_{-0.3}$&$261.5^{+8.5}_{-5.2}$&$[2000.0] $&$1381.9^{+23.0}_{-17.6}$&
\citeauthor{Pello_1991} \citeyear{Pello_1991}\\
DM2&$[-0.9]$&$[-1.3]$&$0.35^{+0.05}_{-0.03}$&$33.3^{+1.2}_{-1.6}$& $25.0^{+1.8}_{-1.1}$&
$750.4^{+100.2}_{-65.5} $&$585.1^{+20.0}_{-9.7}$&\citeauthor{Richard_2010} \citeyear{Richard_2010}\\
BCG1&[46.8] &[12.8] &$0.11^{+0.10}_{-0.01}$&$114.8^{+26.8}_{-31.5} $&[0.05] &
$23.1^{+3.0}_{-1.6}$&$151.9^{+5.9}_{-7.5}$ &Pello et al (in prep)\\
$L*$  Gal &&&&&[0.15] &[45.0] &$185.7^{+5.3}_{-3.3}$\\
\hline
A2667\\
\hline
DM1&$0.2^{+0.5}_{-0.4}$&$1.3^{+0.5}_{-0.4}$ &$0.46^{+0.02}_{-0.02}$ &$-44.4^{+0.2}_{-0.3}$&$79.33^{+1.1}_{-1.1} $&[1298.7]&$1095.0^{+5.0}_{-3.7}$&\citeauthor{Covone_2006} \citeyear{Covone_2006}\\
$L*$ Gal &&&&&[0.15] &[45.0]&$91.3^{+4.5}_{-4.5}$& \citeauthor{Richard_2010} \citeyear{Richard_2010}\\
\hline
A2744\\
\hline
DM1&$-2.1^{+0.3}_{-0.3} $&$1.4^{+0.0}_{-0.4}$ &$0.83^{+0.01}_{-0.02} $&
$90.5^{+1.0}_{-1.1}$ &$85.4^{+5.4}_{-4.5} $&[1000.0] &$607.1^{+7.6}
_{-0.2}$ &\citeauthor{Mahler_2017} \citeyear{Mahler_2017} \\
DM2&$-17.1^{+0.2}_{-0.3} $&$-15.7^{+0.4}_{-0.3}$ &$0.51^{+0.02}_{-0.02}$ &$45.2^{+1.3}_{-0.8}$ &$48.3^{+5.1}_{2.2}$&[1000.0] &$742.8^{+20.1}_{-14.2}$&\citetalias{Richard_2021}\\
BCG1&[0.0] &[0.0] &[0.21]&$[-76.0] $&[0.3] &[28.5] &$355.2^{+11.3}_{-10.2}$&\\
BCG2&$[-17.9]$ &$[-20.0]$ &[0.38]&[14.8] &[0.3] &[29.5] &$321.7^{+15.3}_{-7.3}$&\\
NGal&$[-3.6]$ &[24.7]&[0.72] &$[-33.0]$ &[0.1]&[13.2] &$175.6^{+8.7}_{-13.8}$&\\
SGal&$[-12.7]$ &$[-0.8]$&[0.30] &$[-46.6]$ &[0.1] &$6.8^{+93.3}_{-3.2} $&$10.6^{+43.2}_{-3.6}$&\\
$L*$ Gal &&&&&[0.15] &$13.7^{+1.0}_{0.6} $&$155.5^{+4.2}_{-5.9}$&\\
\hline
A370\\
\hline
DM1&$2.21^{+0.12}_{-0.10} $&$1.33^{+0.05}_{-0.06} $&$0.40^{+0.03}_{-0.03}$ &
$-69.6^{+1.5}_{-1.3}$&$14.7^{+1.0}_{-1.5} $&[800.0] &$394^{+15}_{-9}$&\citeauthor{Lagattuta_2019} \citeyear{Lagattuta_2019}\\
DM2&$2.01^{+0.10}_{-0.23}$&$11.35^{+0.31}_{-0.38}$ &$0.69^{+0.02}_{-0.01} $&
$-122.3^{+0.4}_{-0.6}$ &$137.4^{+0.2}_{-1.3}$&[800.0] &$1039^{+6}_{-14}$& \citetalias{Richard_2021}\\
BCG1&$[-0.01]$&[0.02]&[0.30] &$[-81.9] $&[0.1] &$57.6^{+4.1}_{-5.1}$&$224^{+9}_{-6}$&\\
BCG2&[5.90] &[37.24] &[0.20] &$[-63.9]$ &[0.1] &$77.6^{+6.0}_{-6.8}$&
$388^{+6}_{-9}$&\\
$L* $ Gal&&&&&[0.15]&$10.8^{+0.6}_{-1.2}$&$152^{+2}_{-1}$&\\
\hline 
AS1063\\
\hline
DM1&[0.]&[0.]&0.61&-37.5&130&[1000]&1352&\cite{Beauchesne_2023} \\
DM2&0.0&0.03&0.27&[-35]&0.51&169&328&\\
L* Gal& &&&&[0.15]&[45]&99&\\
\hline
BULLET\\
\hline
DM1&$4.8^{+0.2}_{-0.1}$&$1.2^{+0.5}_{-0.5}$ &$0.68^{+0.03}_{-0.03}$ &
$79.5^{+0.2}_{-0.7}$&$138^{+8}_{-9} $&[1000] &$787^{+19}_{-25}$&
\citetalias{Richard_2021}\\
DM2&$29.9^{+0.0}_{-0.2}$&$26.3^{+0.4}_{-0.5}$ &$0.64^{+0.02}_{-0.02}$&
$55.8^{+0.6}_{-0.9} $&$168^{+4}_{-4} $&[1000] &$1004^{+28}_{-21}$&\\
GAL1&[0.0] &[0.0] &[0.26] &[43.5] &[0]&[150]&$268^{+9}_{-13}$&\\
GAL2&[24.0] &[29.1] &[0.20] &[37.4]&[0]&[112] &$118^{+12}_{-10}$&\\
$L*$ Gal&&&&&[0.15] &$25^{+3}_{-2} $&$165^{+2}_{-3}$&\\
\hline
MACS0257\\
\hline
DM1&$-2.1^{+0.3}_{-0.2} $&$1.8^{+0.3}_{-0.2}$&$0.59^{+0.02}_{-0.02}$&
$200.9^{+1.1}_{-1.2}$ &$62^{+2}_{-2}$ &$1014^{+145}_{-79}$&$877^{+17}_{-16}$&
\citetalias{Richard_2021}\\
DM2&$8.5^{+5.0}_{-2.4} $&$-8.4^{+0.8}_{-2.6}$ &$0.87^{+0.03}_{-0.04}$ &
$150.1^{+2.1}_{-0.8}$ &$189^{+6}_{-12}$ &$1093^{+111}_{-174}$&$733^{+15}_{-22}$&\\
GAL1&$[-14.1]$&[15.1] &$0.39^{+0.13}_{-0.14} $&$6.8^{+35.7}_{-34.5}$&[0] &
$88^{+5}_{-10} $&$184^{+6}_{-5}$&\\
GAL2&$[-10.6] $&[17.6] &[0.50] &$30.6^{+8.2}_{-12.5} $&[0] &[40]&$171^{+22}_{-16} $& \\
$L*$ Gal&&&&&[0.15] &$51^{+4}_{-3}$&$154^{+3}_{-2}$&\\
\hline
MACS0329\\
\hline
DM1&$0.4^{+0.1}_{-0.2}$ &$-0.4^{+0.2}_{-0.1}$ &$0.16^{+0.02}_{-0.01}$&
$64.7^{+4.1}_{-2.1}$&$55^{+3}_{-4}$&[1000] &$959^{+14}_{-17}$&\citetalias{Richard_2021}\\
DM2&$43.2^{+1.5}_{-0.7}$&$17.8^{+1.2}_{-1.5}$&[0.30] &$73.1^{+5.0}_{-4.4} $&
$7^{+10}_{-25}$&[1000]&$552^{+30}_{-40}$&\\
GAL1&[0.0]&[0.0]&[0.19] &$[-73.6]$ &[0] &[98] &$281^{+15}_{-30}$&\\
GAL2&$[-12.7] $&$[-39.9] $&[0.14]&[56.9] &[0] &[41]&$218^{+4}_{-4}$&\\
$L*$ Gal  &&&&&[0.15]&[45] &$159^{+4}_{-4}$&\\
  \hline        
     MACS0416\\
\hline
DM1 &$-2.9^{+0.3}_{-0.3} $&$1.4^{+0.3}_{-0.2} $&$0.78^{+0.01}_{-0.01}$ &
$142.1^{+0.4}_{-0.4}  $ &$59^{+2}_{-2}  $&[1000]   &$731^{+10}_{-11}  $ &\citetalias{Richard_2021} \\
DM2&$22.6^{+0.3}_{-0.2}$&$-42.4^{+0.4}_{-0.6} $&$0.69^{+0.01}_{-0.01} $&
$127.1^{+0.2}_{-0.3}  $&$92^{+2}_{-3}  $&[1000] &$940^{+12}_{-11}$& \\
GAL1&[31.8] &$[-65.5]$ &[0.04] &$[-40.4]$ &[0] &[62] &$137^{+10}_{-11}  $&\\
GAL2 &$-37.2^{+0.6}_{-0.8} $&$7.8^{+1.3}_{-1.1} $ &$0.82^{+0.03}_{-0.03}  $ &
$118.5^{+3.9}_{-3.8}  $ &[25] &[200] &$252^{+10}_{-7}  $ &\\
$L*$ Gal  &&&&&[0.15] &$36^{+3}_{-2} $&$137^{+2}_{-2}$&\\
\hline
MACS0451\\
\hline
DM1&4.8&1.0&0.8&-8.1&88&[1000]&763& Basto et al. (in prep)\\
DM2&4.0&7.3&0.78&27.8&103&[1000]&933&\\
GAL1&[-57]&[-7.7]&[0.22]&-12.74&[50]&[1000]&253&\\
L* Gal&&&&&[0.15]&[10]&99.5&\\

\hline
MACS0520\\
\hline
DM1&2.2&1.0&0.38&7.0&78&[1000]&1186&  Basto et al. (in prep) \\
DM2&[-0.]&[0.]&0.19&[-3.1]&0.1&6&597& \\
L* Gal&&&&&0.15&66.&329&\\

\hline
       MACS0940\\
       \hline
       
DM1&$0.6^{+0.8}_{-0.7} $&$0.6^{+1.4}_{-0.2}$&$0.46^{+0.19}_{-0.04}$ &
$23.5^{+2.0}_{-1.2}$&$31^{+79}_{-8}$&$1386^{+565}_{-70}$ &$496^{+223}_{-42}$&
\citetalias{Richard_2021}\\
GAL1&$[-0.1] $&[0.1] &$0.37^{+0.09}_{-0.06}  $&$[-7.7]$&[0]&[52]&
$436^{+15}_{-30}$&\\
GAL2  &$[-11.8] $&[3.1] &$0.66^{+0.07}_{-0.29}  $&$5.9^{+19.9}_{-20.9}  $&[0] &
[17] &$195^{+17}_{-14}$&\\
$L*$ Gal  &&&&&[0.15] &$62^{+62}_{-94} $&$162^{+28}_{-7}$&\\
\hline
       MACS1206\\
\hline
DM1&$-0.1^{+0.0}_{-0.0}$&$0.7^{+0.0}_{-0.0} $&$0.63^{+0.01}_{-0.01}$ &
$19.9^{+0.2}_{-0.1}$&$44^{+0}_{-1} $&[1000] &$888^{+7}_{-6}$&\citetalias{Richard_2021}
\\
DM2&$9.5^{+0.5}_{-0.2}$&$5.7^{+0.4}_{-0.3}$&$0.70^{+0.02}_{-0.03} $&
$114.7^{+0.7}_{-0.5}$&$94^{+3}_{-3} $&[1000] &$575^{+6}_{-9}$&\\
GAL1&$[-0.1]$&[0.0]&[0.71]&[14.4] &$1^{+0}_{-1} $&$20^{+1}_{-1} $&$355^{+11}_{-6}$&
\\
GAL2&[35.8]&[16.1]&[0.23] &$133.8^{+69.7}_{-47.6} $&[0] &$4^{+1}_{-1} $&
$275^{+18}_{-11}$&\\
$L*$ Gal&&&&&$34^{+5}_{-1} $&$198^{+5}_{-6}$&\\
\hline
        MACS2214\\
        \hline
DM1&$-1.2^{+0.1}_{-0.2} $&$0.7^{+0.1}_{-0.1}$&$0.55^{+0.01}_{-0.01} $&$147.5^{+0.7}_{-0.8} $&
$38^{+1}_{-1}$&[1000]&$903^{+15}_{-18}$&\citetalias{Richard_2021}\\
DM2&$-20.8^{+0.2}_{-0.2}$&$17.2^{+0.3}_{-0.2}$&$0.70^{+0.06}_{-0.05} $&$112.0^{+3.2}_{-3.5}$&
$20^{+8}_{-5} $&[1000]&$299^{+11}_{-34}$&\\
GAL1&[0.0] &[0.0] &[0.20] &$151.1^{+34.0}_{-45.5} $&$4^{+14}_{-3}$&$9^{+16}_{-12} $&$79^{+61}_{-28}$&\\
GAL2&[8.2]&[18.8]&$0.48^{+0.07}_{-0.04} $&[0.0] &[0] &$81^{+2}_{-3} $&$169^{+5}_{-9}$&\\
$L*$ Gal &&&&&[0.15] &$46^{+3}_{-0}$&$111^{+2}_{-3}$&\\
\hline
RXJ1347\\
\hline
DM1&$0.4^{+0.1}_{-0.1} $&$5.1^{+0.4}_{-0.2} $&$0.76^{+0.02}_{-0.02} $&$111.8^{+0.5}_{-0.7} $&
$37^{+1}_{-1} $&[1000]   &$638^{+24}_{-24}  $&\citetalias{Richard_2021}\\
DM2&$-13.6^{+0.2}_{-0.1}$&$-4.5^{+0.2}_{-0.4} $&$0.78^{+0.00}_{-0.01}$ &$121.4^{+0.2}_{-0.1} $&
$78^{+2}_{-2} $&[1000]&$850^{+8}_{-4}$&\\
GAL1&[0.0]&$[-0.0]$&[0.23]&$[-86.9]$&[0]&$84^{+13}_{-13}$ &$354^{+7}_{-5}$&\\
GAL2&$[-17.8]$ &$[-2.1] $&[0.30] &$[-64.1] $&[0] &$94^{+6}_{-4} $&$364^{+2}_{-3}$&\\
$L*$ Gal&&&&&[0.15]&$81^{+9}_{-9} $&$135^{+3}_{-4}$&\\
\hline
SMACS2031\\
\hline
DM1 &$0.4^{+0.1}_{-0.1} $&$-0.7^{+0.1}_{-0.1}$&$0.31^{+0.02}_{-0.02}$&$4.4^{+1.0}_{-1.1} $&
$29^{+1}_{-1}$&$[1000] $&$624^{+12}_{-11}$&\citetalias{Richard_2021}\\
DM2&$61.2^{+0.4}_{-0.5}$&$25.3^{+0.4}_{-0.3}$&$0.46^{+0.05}_{-0.04} $&$6.4^{+2.1}_{-3.9}$&
$112^{+8}_{-7}$&[1000] &$1037^{+20}_{-19}$&\\
GAL1&[0.1] &$[-0.1] $&[0.09] &$[-0.4]$ &[0.0] &$128^{+12}_{-46}$&$242^{+3}_{-4}$&\\
$L*$ Gal&&&&&[0.15] &$9^{+2}_{-1}$&$161^{+19}_{-9}$&\\
\hline
SMACS2131\\
\hline
DM1&$-3.2^{+0.4}_{-0.2} $&$3.2^{+0.1}_{-0.3} $&$0.66^{+0.01}_{-0.01}$&$155.2^{+0.2}_{-0.3}$&
$84^{+1}_{-1} $&[1000] &$866^{+15}_{-5}$&\citetalias{Richard_2021} \\
DM2&$21.2^{+0.9}_{-0.3}$&[17.0] &$0.53^{+0.02}_{-0.02}$ &$82.7^{+6.4}_{-6.9}$&
$95^{+14}_{-6} $&[1000] &$452^{+44}_{-12}$&\\
GAL1&$[-0.0]$ &[0.0] &[0.11] &$59.0^{+1.4}_{-1.3}$ &[0] &$155^{+3}_{-14} $&$399^{+0}_{-1}$&\\
GAL2&[6.7] &$[-2.2] $&[0.76] &$7.2^{+9.7}_{-6.4}$ &[0] &$96^{+8}_{-17} $&$93^{+18}_{-12}$&\\
$L*$ Gal&&&&&[0.15] &$93^{+6}_{-21} $&$202^{+4}_{-1}$ & \\

         \hline
           \hline
  
   \end{longtable}

\end{appendix}

\end{document}